\documentclass[12pt]{iopart}
\usepackage{iopams}
\usepackage{setstack}
\usepackage{amssymb}
\usepackage{latexsym}
\usepackage{color}
\usepackage{graphicx}
     
\begin{document}
\title[Properties of finite Gaussians and the discrete-continuous transition]{Properties of finite Gaussians and\\ the discrete-continuous transition}
\author{Nicolae Cotfas and Daniela Dragoman}
\address{University of Bucharest,  Physics Department,\\ P.O. Box MG-11, 077125 Bucharest, Romania}

\eads{\mailto{ncotfas@yahoo.com}, \mailto{danieladragoman@yahoo.com}}

\begin{abstract}
Weyl's formulation of quantum mechanics opened the possibility of studying the dynamics 
of quantum systems both in infinite-dimensional and finite-dimensional systems. 
Based on Weyl's approach, generalized by Schwinger, a self-consistent theoretical framework describing physical systems characterised by a finite-dimensional  space of states has been created. The  used mathematical formalism is further developed by adding finite-dimensional versions of some notions and results from the continuous case. 
Discrete versions of the continuous Gaussian functions have been defined by using the Jacobi theta functions. We continue the investigation of the properties of these finite Gaussians by following the analogy with the continuous case. We study the uncertainty relation of finite Gaussian states, the form of the associated Wigner quasi-distribution and the evolution under free-particle and quantum harmonic oscillator Hamiltonians. In all cases, a particular emphasis is put on the recovery of the known continuous-limit results when the dimension $d$ of the system increases. 
\end{abstract}
\maketitle

\section{Introduction}
\bigskip

The continuous Gaussian functions 
\[
g_\kappa :\mathbb{R}\longrightarrow \mathbb{R},\qquad g_\kappa (x)=\mathrm{e}^{-\frac{\kappa }{2}x^2}\qquad \qquad \qquad \kappa \!\in \!(0,\infty )
\]
play a fundamental role in physics, particularly in quantum mechanics, due to their remarkable properties, among which we mention:  
\begin{enumerate}
\item With the Fourier transform defined as 
\[
\mathcal{F}[f](\xi )=\frac{1}{\sqrt{2\pi }}\int_{-\infty }^\infty {\rm e}^{{\rm i}\xi x}f(x)\, dx
\]
we have
\begin{equation}\label{gauss-cont1}
\mathcal{F}[g_\kappa ]=\frac{1}{\sqrt{\kappa }}\, g_{\frac{1}{\kappa }}.
\end{equation}
\item The function $g_1 $ is an eigenfunction of a second-order differential operator
\begin{equation}\label{gauss-cont2}
\left( -\frac{d^2}{dx^2}+x^2 \right){\rm e}^{-\frac{1}{2}x^2}= {\rm e}^{-\frac{1}{2}x^2}.
\end{equation}
\item The function $g_\kappa  $ is a minimum uncertainty state for the coordinate-momentum
\begin{equation}\label{gauss-cont3}
\Delta \hat x\, \Delta \hat p=\frac{1}{2}
\end{equation}
\item  The Wigner quasi-distribution \cite{wigner} corresponding to $g_\kappa $ defined as
\[
W(x,p) =\frac{1}{\pi}\int_{-\infty}^\infty du\, \mathrm{e}^{2{\rm i}pu }g_\kappa \left(x-u\right)\, g_\kappa ^*\left(x+u\right) 
\]
is, up to a multiplicative constant, a product of two Gaussian functions
\begin{equation}\label{gauss-cont4}
W(x,p)=2\sqrt{\frac{\pi }{\kappa }}\,  \mathrm{e}^{-\kappa  x^2} \, \mathrm{e}^{-\frac{1}{\kappa }p^2}.
\end{equation}
\end{enumerate}

\begin{center}
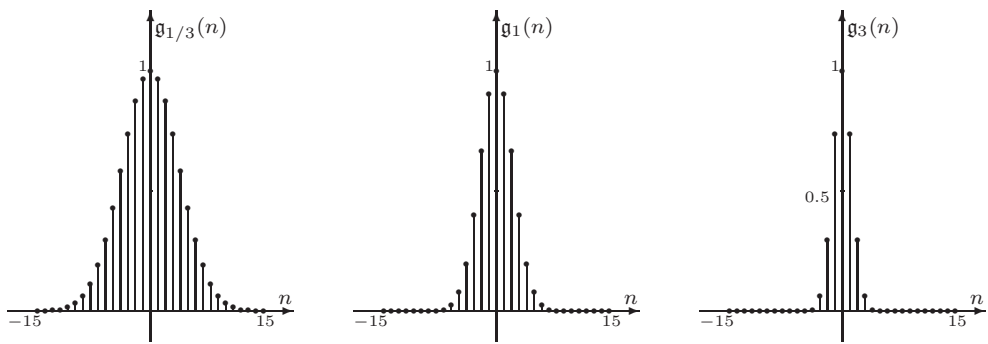
\begin{figure}[h]
\setlength{\unitlength}{2mm}
\begin{picture}(70,24)(-10,0)
\put(0,2){\vector(1,0){19}}
\put(23,2){\vector(1,0){19}}
\put(46,2){\vector(1,0){19}}
\put(9.5,0){\vector(0,1){22}}
\put(32.5,0){\vector(0,1){22}}
\put(55.5,0){\vector(0,1){22}}

\put(    2.00000,   2.01081){\circle*{0.3}}
\put(    2.50000,   2.02223){\circle*{0.3}}
\put(    3.00000,   2.05334){\circle*{0.3}}
\put(    3.50000,   2.12354){\circle*{0.3}}
\put(    4.00000,   2.26853){\circle*{0.3}}
\put(    4.50000,   2.54582){\circle*{0.3}}
\put(    5.00000,   3.03703){\circle*{0.3}}
\put(    5.50000,   3.84160){\circle*{0.3}}
\put(    6.00000,   5.05672){\circle*{0.3}}
\put(    6.50000,   6.74214){\circle*{0.3}}
\put(    7.00000,   8.87626){\circle*{0.3}}
\put(    7.50000,  11.31941){\circle*{0.3}}
\put(    8.00000,  13.80547){\circle*{0.3}}
\put(    8.50000,  15.97774){\circle*{0.3}}
\put(    9.00000,  17.46854){\circle*{0.3}}
\put(    9.50000,  18.00000){\circle*{0.3}}
\put(   10.00000,  17.46854){\circle*{0.3}}
\put(   10.50000,  15.97774){\circle*{0.3}}
\put(   11.00000,  13.80547){\circle*{0.3}}
\put(   11.50000,  11.31941){\circle*{0.3}}
\put(   12.00000,   8.87626){\circle*{0.3}}
\put(   12.50000,   6.74214){\circle*{0.3}}
\put(   13.00000,   5.05672){\circle*{0.3}}
\put(   13.50000,   3.84160){\circle*{0.3}}
\put(   14.00000,   3.03703){\circle*{0.3}}
\put(   14.50000,   2.54582){\circle*{0.3}}
\put(   15.00000,   2.26853){\circle*{0.3}}
\put(   15.50000,   2.12354){\circle*{0.3}}
\put(   16.00000,   2.05334){\circle*{0.3}}
\put(   16.50000,   2.02223){\circle*{0.3}}
\put(   17.00000,   2.01081){\circle*{0.3}}
\put(    2.00000,2){\line(0,1){    .01081}}
\put(    2.50000,2){\line(0,1){    .02223}}
\put(    3.00000,2){\line(0,1){    .05334}}
\put(    3.50000,2){\line(0,1){    .12354}}
\put(    4.00000,2){\line(0,1){    .26853}}
\put(    4.50000,2){\line(0,1){    .54582}}
\put(    5.00000,2){\line(0,1){   1.03703}}
\put(    5.50000,2){\line(0,1){   1.84160}}
\put(    6.00000,2){\line(0,1){   3.05672}}
\put(    6.50000,2){\line(0,1){   4.74214}}
\put(    7.00000,2){\line(0,1){   6.87626}}
\put(    7.50000,2){\line(0,1){   9.31941}}
\put(    8.00000,2){\line(0,1){  11.80547}}
\put(    8.50000,2){\line(0,1){  13.97774}}
\put(    9.00000,2){\line(0,1){  15.46854}}
\put(    9.50000,2){\line(0,1){  16.00000}}
\put(   10.00000,2){\line(0,1){  15.46854}}
\put(   10.50000,2){\line(0,1){  13.97774}}
\put(   11.00000,2){\line(0,1){  11.80547}}
\put(   11.50000,2){\line(0,1){   9.31941}}
\put(   12.00000,2){\line(0,1){   6.87626}}
\put(   12.50000,2){\line(0,1){   4.74214}}
\put(   13.00000,2){\line(0,1){   3.05672}}
\put(   13.50000,2){\line(0,1){   1.84160}}
\put(   14.00000,2){\line(0,1){   1.03703}}
\put(   14.50000,2){\line(0,1){    .54582}}
\put(   15.00000,2){\line(0,1){    .26853}}
\put(   15.50000,2){\line(0,1){    .12354}}
\put(   16.00000,2){\line(0,1){    .05334}}
\put(   16.50000,2){\line(0,1){    .02223}}
\put(   17.00000,2){\line(0,1){    .01081}}
\put(   25.00000,   2.00000){\circle*{0.3}}
\put(   25.50000,   2.00000){\circle*{0.3}}
\put(   26.00000,   2.00000){\circle*{0.3}}
\put(   26.50000,   2.00000){\circle*{0.3}}
\put(   27.00000,   2.00000){\circle*{0.3}}
\put(   27.50000,   2.00064){\circle*{0.3}}
\put(   28.00000,   2.00436){\circle*{0.3}}
\put(   28.50000,   2.02440){\circle*{0.3}}
\put(   29.00000,   2.11156){\circle*{0.3}}
\put(   29.50000,   2.41656){\circle*{0.3}}
\put(   30.00000,   3.27004){\circle*{0.3}}
\put(   30.50000,   5.16174){\circle*{0.3}}
\put(   31.00000,   8.42704){\circle*{0.3}}
\put(   31.50000,  12.66771){\circle*{0.3}}
\put(   32.00000,  16.45798){\circle*{0.3}}
\put(   32.50000,  18.00000){\circle*{0.3}}
\put(   33.00000,  16.45798){\circle*{0.3}}
\put(   33.50000,  12.66771){\circle*{0.3}}
\put(   34.00000,   8.42704){\circle*{0.3}}
\put(   34.50000,   5.16174){\circle*{0.3}}
\put(   35.00000,   3.27004){\circle*{0.3}}
\put(   35.50000,   2.41656){\circle*{0.3}}
\put(   36.00000,   2.11156){\circle*{0.3}}
\put(   36.50000,   2.02440){\circle*{0.3}}
\put(   37.00000,   2.00436){\circle*{0.3}}
\put(   37.50000,   2.00064){\circle*{0.3}}
\put(   38.00000,   2.00000){\circle*{0.3}}
\put(   38.50000,   2.00000){\circle*{0.3}}
\put(   39.00000,   2.00000){\circle*{0.3}}
\put(   39.50000,   2.00000){\circle*{0.3}}
\put(   40.00000,   2.00000){\circle*{0.3}}
\put(   25.00000,2){\line(0,1){    .00000}}
\put(   25.50000,2){\line(0,1){    .00000}}
\put(   26.00000,2){\line(0,1){    .00000}}
\put(   26.50000,2){\line(0,1){    .00000}}
\put(   27.00000,2){\line(0,1){    .00000}}
\put(   27.50000,2){\line(0,1){    .00064}}
\put(   28.00000,2){\line(0,1){    .00436}}
\put(   28.50000,2){\line(0,1){    .02440}}
\put(   29.00000,2){\line(0,1){    .11156}}
\put(   29.50000,2){\line(0,1){    .41656}}
\put(   30.00000,2){\line(0,1){   1.27004}}
\put(   30.50000,2){\line(0,1){   3.16174}}
\put(   31.00000,2){\line(0,1){   6.42704}}
\put(   31.50000,2){\line(0,1){  10.66771}}
\put(   32.00000,2){\line(0,1){  14.45798}}
\put(   32.50000,2){\line(0,1){  16.00000}}
\put(   33.00000,2){\line(0,1){  14.45798}}
\put(   33.50000,2){\line(0,1){  10.66771}}
\put(   34.00000,2){\line(0,1){   6.42704}}
\put(   34.50000,2){\line(0,1){   3.16174}}
\put(   35.00000,2){\line(0,1){   1.27004}}
\put(   35.50000,2){\line(0,1){    .41656}}
\put(   36.00000,2){\line(0,1){    .11156}}
\put(   36.50000,2){\line(0,1){    .02440}}
\put(   37.00000,2){\line(0,1){    .00436}}
\put(   37.50000,2){\line(0,1){    .00064}}
\put(   38.00000,2){\line(0,1){    .00000}}
\put(   38.50000,2){\line(0,1){    .00000}}
\put(   39.00000,2){\line(0,1){    .00000}}
\put(   39.50000,2){\line(0,1){    .00000}}
\put(   40.00000,2){\line(0,1){    .00000}}
\put(   48.00000,   2.00000){\circle*{0.3}}
\put(   48.50000,   2.00000){\circle*{0.3}}
\put(   49.00000,   2.00000){\circle*{0.3}}
\put(   49.50000,   2.00000){\circle*{0.3}}
\put(   50.00000,   2.00000){\circle*{0.3}}
\put(   50.50000,   2.00000){\circle*{0.3}}
\put(   51.00000,   2.00000){\circle*{0.3}}
\put(   51.50000,   2.00000){\circle*{0.3}}
\put(   52.00000,   2.00000){\circle*{0.3}}
\put(   52.50000,   2.00028){\circle*{0.3}}
\put(   53.00000,   2.00800){\circle*{0.3}}
\put(   53.50000,   2.12346){\circle*{0.3}}
\put(   54.00000,   3.03703){\circle*{0.3}}
\put(   54.50000,   6.74214){\circle*{0.3}}
\put(   55.00000,  13.80547){\circle*{0.3}}
\put(   55.50000,  18.00000){\circle*{0.3}}
\put(   56.00000,  13.80547){\circle*{0.3}}
\put(   56.50000,   6.74214){\circle*{0.3}}
\put(   57.00000,   3.03703){\circle*{0.3}}
\put(   57.50000,   2.12346){\circle*{0.3}}
\put(   58.00000,   2.00800){\circle*{0.3}}
\put(   58.50000,   2.00028){\circle*{0.3}}
\put(   59.00000,   2.00000){\circle*{0.3}}
\put(   59.50000,   2.00000){\circle*{0.3}}
\put(   60.00000,   2.00000){\circle*{0.3}}
\put(   60.50000,   2.00000){\circle*{0.3}}
\put(   61.00000,   2.00000){\circle*{0.3}}
\put(   61.50000,   2.00000){\circle*{0.3}}
\put(   62.00000,   2.00000){\circle*{0.3}}
\put(   62.50000,   2.00000){\circle*{0.3}}
\put(   63.00000,   2.00000){\circle*{0.3}}
\put(   48.00000,2){\line(0,1){    .00000}}
\put(   48.50000,2){\line(0,1){    .00000}}
\put(   49.00000,2){\line(0,1){    .00000}}
\put(   49.50000,2){\line(0,1){    .00000}}
\put(   50.00000,2){\line(0,1){    .00000}}
\put(   50.50000,2){\line(0,1){    .00000}}
\put(   51.00000,2){\line(0,1){    .00000}}
\put(   51.50000,2){\line(0,1){    .00000}}
\put(   52.00000,2){\line(0,1){    .00000}}
\put(   52.50000,2){\line(0,1){    .00028}}
\put(   53.00000,2){\line(0,1){    .00800}}
\put(   53.50000,2){\line(0,1){    .12346}}
\put(   54.00000,2){\line(0,1){   1.03703}}
\put(   54.50000,2){\line(0,1){   4.74214}}
\put(   55.00000,2){\line(0,1){  11.80547}}
\put(   55.50000,2){\line(0,1){  16.00000}}
\put(   56.00000,2){\line(0,1){  11.80547}}
\put(   56.50000,2){\line(0,1){   4.74214}}
\put(   57.00000,2){\line(0,1){   1.03703}}
\put(   57.50000,2){\line(0,1){    .12346}}
\put(   58.00000,2){\line(0,1){    .00800}}
\put(   58.50000,2){\line(0,1){    .00028}}
\put(   59.00000,2){\line(0,1){    .00000}}
\put(   59.50000,2){\line(0,1){    .00000}}
\put(   60.00000,2){\line(0,1){    .00000}}
\put(   60.50000,2){\line(0,1){    .00000}}
\put(   61.00000,2){\line(0,1){    .00000}}
\put(   61.50000,2){\line(0,1){    .00000}}
\put(   62.00000,2){\line(0,1){    .00000}}
\put(   62.50000,2){\line(0,1){    .00000}}
\put(   63.00000,2){\line(0,1){    .00000}}

\put(0,1){$\scriptscriptstyle{-15}$}
\put(16.5,1){$\scriptscriptstyle{15}$}
\put(23,1){$\scriptscriptstyle{-15}$}
\put(39.5,1){$\scriptscriptstyle{15}$}
\put(46,1){$\scriptscriptstyle{-15}$}
\put(62.5,1){$\scriptscriptstyle{15}$}

\put(8.7,18){$\scriptscriptstyle{1}$}
%\put(7.8,9.55){$\scriptscriptstyle{0.5}$}
\put(9.4, 10){\line(1,0){0.2}}
\put(18,2.5){$\scriptstyle{n}$}
\put(9.8,20.5){$\scriptstyle{\mathfrak{g}_{1/3}(n)}$}

\put(31.7,18){$\scriptscriptstyle{1}$}
%\put(30.8,9.55){$\scriptscriptstyle{0.5}$}
\put(32.4, 10){\line(1,0){0.2}}
\put(41,2.5){$\scriptstyle{n}$}
\put(32.8,20.5){$\scriptstyle{\mathfrak{g}_{1}(n)}$}

\put(54.7,18){$\scriptscriptstyle{1}$}
\put(53,9.3){$\scriptscriptstyle{0.5}$}
\put(55.4, 10){\line(1,0){0.2}}
\put(64,2.5){$\scriptstyle{n}$}
\put(55.8,20.5){$\scriptstyle{\mathfrak{g}_{3}(n)}$}
\end{picture}
\caption{The functions $\mathfrak{g}_{1/3}$ (left),  $\mathfrak{g}_1$ (center) and $\mathfrak{g}_{3}$ (right) in the case $d=31$.}\label{gaussians}
\end{figure}
\end{center}

Let $d=2s+1\in \{ 3,5,7,...\}$ be a fixed odd positive integer, and let $\mathbb{Z}_d\!=\!\mathbb{Z}/d\mathbb{Z}$ be the ring of integers modulo $d$ for which we use $\{-s,-s\!+\!1,...,s\!-\!1,s\}$ as a set of `standard' representatives. 
The number $d$ represents the dimension of the Hilbert space describing the states of the investigated quantum systems.

The Jacobi theta function \cite{Magnus,Vilenkin,Whittaker}
\[
\theta _3(z,\tau )=\sum_{\alpha =-\infty }^\infty {\rm e}^{{\rm i}\pi \tau \alpha ^2}\, {\rm e}^{2\pi {\rm i}\alpha z}
\]
has several remarkable properties among which we mention:
\[
\theta _3(z+m+n\tau ,\tau )={\rm e}^{-{\rm i}\pi \tau n^2}\, {\rm e}^{-2\pi {\rm i}nz}\, \theta _3(z,\tau )
\]
\[
\theta _3(z,{\rm i}\tau )=\frac{1}{\sqrt{\tau }}\, {\rm exp}^{-\frac{\pi z^2}{\tau }}\, \theta _3 \left( \frac{z}{{\rm i}\tau },\frac{\rm i}{\tau }\right)
\]
and \cite{Ruzzi}
\begin{equation} \label{Ruzzi}
\theta _3 \left( \frac{k}{d},\frac{{\rm i}\kappa }{d}\right)=\frac{1}{\sqrt{\kappa d}}\, \sum_{n=-s}^{s}{\rm e}^{-\frac{2\pi {\rm i}}{d}kn}\, \theta _3 \left( \frac{n}{d},\frac{\rm i}{ \kappa d}\right).
\end{equation}
For any $\kappa \in (0,\infty )$ the function 
\[
\mathfrak{g}_\kappa  :\mathbb{Z}_d\longrightarrow \mathbb{R},\qquad \mathfrak{g}_\kappa  (n)=\sum_{\alpha =-\infty }^\infty \mathrm{e}^{-\frac{\kappa \pi }{d} (\alpha d +n)^2}
\]
can be expressed in terms of the Jacobi function $\theta _3$ as
\[
\mathfrak{g}_\kappa (n)=\frac{1}{\sqrt{\kappa d}}\, \theta _3\left( \frac{n}{d},\frac{{\rm i}}{\kappa d} \right)
\]
and by using the finite Fourier transform 
\[
F[\psi ] (k)=\frac{1}{\sqrt{d}}\sum_{n=-s}^{s}{\rm e}^{\frac{2\pi {\rm i}}{d}kn}\psi (n)
\]
Ruzzi's relation (\ref{Ruzzi}) can be written in a form identical to (\ref{gauss-cont1}), namely,
\begin{equation} \label{Fouriergk}
F[\mathfrak{g} _\kappa ]=\frac{1}{\sqrt{\kappa }}\, \mathfrak{g} _{\frac{1}{\kappa }}.
\end{equation}

This property of the function $\mathfrak{g}_\kappa $ and the shape of its graph (see figure \ref{gaussians}) show that it can be regarded as a finite version of the Gaussian function $g_\kappa $. We call $\mathfrak{g}_\kappa $ a {\em finite Gaussian}, and our main purpose is to prove the existence of a  finite version for the relations (1)-(4). We recover well-known results for continuous Gaussians for large $d$ values, i.e. at the discrete-continuous transition, and emphasize conditions under which the evolution of continuous Gaussian states can be retrieved from the evolution of their discrete counterparts.

The finite Gaussians $\mathfrak{g}_\kappa $ represent a generalization of Mehta's function $f_0$, and the relation  (\ref{Fouriergk}) is a generalization of Mehta's relation $F[f_0]= f_0$. More precisely, Mehta has proved  \cite{Mehta} that the functions 
\[
\begin{array}{l}
f_k(n)=\sum\limits_{\alpha =-\infty }^\infty \mathrm{e}^{-\frac{\pi }{d} (\alpha d +n)^2}\, H_k\left( \sqrt{\frac{2\pi }{d}}(\alpha d+n) \right)
\end{array}
\]
defined by using the Hermite polynomials are eigenvectors of the finite Fourier transform  
\[
F[f_k]={\rm i}^k\, f_k\qquad {\rm for\ any}\quad  k\!\in \!\{ 0,1,2,...\}.
\]
Since $H_0(x)=1$, we have $f_0=\mathfrak{g}_1$ and the relation $F[f_0]= f_0$ coincides to $F[\mathfrak{g}_1 ]=\mathfrak{g}_1$. 
Mehta's results are mainly based on the following remarks, which we will use in the following:
\begin{itemize}
\item [a)]  A periodic function with period $d$, namely,
\[
\begin{array}{l}
\Phi :\mathbb{Z}\longrightarrow \mathbb{R}, \qquad \Phi (n)=\sum\limits_{\alpha =-\infty }^\infty  \varphi \left( \sqrt{\frac{2\pi }{d}}(\alpha d+n) \right)
\end{array}
\]
can be defined by starting from any function $\varphi :\mathbb{R}\longrightarrow \mathbb{R}$ for which the series is absolutely convergent. This is similar to a Zak \cite{Zak} or Weil \cite{Weil} transform.
\item[b)] The relation 
\[
\sum_{\alpha =-\infty }^\infty \int_{\alpha \sqrt{2\pi d}}^{(\alpha +1)\sqrt{2\pi d}}\varphi (x)\, dx=\int_{-\infty }^\infty \varphi (x)\, dx
\]
is true  for any function $\varphi :\mathbb{R}\longrightarrow \mathbb{R}$ for which the integral is convergent.
\item[c)] The relation 
\[
\sum_{\alpha =-\infty }^\infty  \varphi (\alpha )=\sum_{n=-s}^{s}\sum_{\alpha =-\infty }^\infty  \varphi (\alpha d+n)
\]
is satisfied for any function $\varphi :\mathbb{Z}\longrightarrow \mathbb{R}$ for which the series is absolutely convergent.
\end{itemize}

In section \ref{Fqs} we review some elements of the mathematical formalism used in the case of the quantum systems with finite-dimensional Hilbert space, in a form suitable for our purpose. We show that in the case of the free evolution, the time dependent state is periodic in time. 
The behaviour of the commutator of the position and momentum operators in the limit of large $d$ is investigated in section \ref{Comm}. Ruzzi has obtained the relation (\ref{Fouriergk}) by using the properties of  $\theta $-functions. In  section \ref{Mus} we present an elementary proof based on a)-c) for this finite version of relation (\ref{gauss-cont1}), and show that $\mathfrak{g}_1$ is almost a minimum uncertainty state.
In  section \ref{Sot} we investigate numerically the quantum oscillator Hamiltonian.
The tendency to have equidistant energy levels becomes evident only for $d$ large enough.
In section \ref{Or} we show, for the first time to our knowledge, that the existence of commensurate or equidistant energy levels is a sufficient condition for the occurrence of revivals. 

In  section \ref{DWF}, by using a)-c) as mathematical tools, we prove that  the  Wigner function corresponding to a finite Gaussian can be written as a sum of four products of finite Gaussians. The obtained formula is our main result, and can be regarded as a finite version of the relation (\ref{gauss-cont4}). In the particular case $\kappa =1$, an expression of the  Wigner function corresponding to a finite Gaussian $\mathfrak{g}_1$ has been previously obtained by Marchiolli and Ruzzi \cite{marchioli}. They proved that, up to a multiplicative constant, the  Wigner function  corresponding to
\[
\mathfrak{g}_1 (n)=\frac{1}{\sqrt{ d}}\, \theta _3\left( \frac{n}{d},\frac{{\rm i}}{ d} \right)
\]
is
\[
W(n,m)=\theta _3\left( \frac{n}{d},\frac{{\rm i}}{ 2d} \right)\, \theta _3\left( \frac{2m}{d},\frac{2{\rm i}}{ d} \right) +
\theta _4\left( \frac{n}{d},\frac{{\rm i}}{ 2d} \right)\, \theta _2\left( \frac{2m}{d},\frac{2{\rm i}}{ d} \right) 
\]
where $\theta _2$ and $\theta _4$ are the Jacobi functions
\[
\theta _2(z,\tau )=\sum_{\alpha =-\infty }^\infty {\rm e}^{{\rm i}\pi \tau \left(\alpha +\frac{1}{2}\right)^2}\, {\rm e}^{2\pi {\rm i}\left(\alpha +\frac{1}{2}\right)z}
\]
and
\[
\theta _4(z,\tau )=\sum_{\alpha =-\infty }^\infty (-1)^\alpha {\rm e}^{{\rm i}\pi \tau \alpha ^2}\, {\rm e}^{2\pi {\rm i}\alpha z}.
\]

\bigskip
%
%%%%%%%%%%%%%%%%%%%%%%%%%%%%%%%
\section{Quantum systems with finite-dimensional Hilbert space} \label{Fqs}
In the case of a quantum particle moving along a straight line, the possible positions form the set $\mathbb{R}=(-\infty ,\infty )$, and the space describing the states of the system is the infinite-dimensional Hilbert space $L^2(\mathbb{R})$ of all the square integrable functions $\psi :\mathbb{R}\longrightarrow \mathbb{C}$.

 We obtain a very simplified version of this continuous
one-dimensional system by assuming that we can distinguish only a finite number $d$ of positions for our particle. In this simplified version, the space describing the states of the system is the $d$-dimensional Hilbert space $\mathbb{C}^d$. 
Assuming that $d$ is an odd number, $d\!=\!2s\!+\!1$,  the space $\mathbb{C}^d$ can be identified with the space $\mathcal{H}$ of all the functions
\[
\psi :\{-s,-s\!+\!1,...,s\!-\!1,s\}\longrightarrow \mathbb{C}
\]
by using the one-to-one mapping
\[
\mathcal{H}\longrightarrow \mathbb{C}^d:\psi \mapsto (\psi (-s),\psi (-s+1),...,\psi (s-1),\psi (s)).
\]
We choose an orthonormal basis $\{ |n\rangle \}_{n\in \mathbb{Z}_d}$ in $\mathcal{H}$  and define the `position' operator
\begin{equation} \label{defq}
\begin{array}{l}
Q:\mathcal{H}\longrightarrow \mathcal{H},\qquad Q=\sqrt{\frac{2\pi }{d}}\sum\limits_{n=-s}^{s}n\, |n\rangle \langle n|.
\end{array}
\end{equation}
The finite Fourier transform
\begin{equation}
F:\mathcal{H}\longrightarrow \mathcal{H},\qquad F=\frac{1}{\sqrt{d}}\sum_{n,n'=-s}^{s}{\rm e}^{\frac{2\pi {\rm i}}{d}nn'}|n\rangle \langle n'|
\end{equation}
allows us to consider a second orthonormal basis 
$\{ |\tilde k\rangle \}_{k\in \mathbb{Z}_d}$, where
\[
|\tilde k\rangle  \!=\!F|k\rangle =F^+|-k\rangle \!=\!\frac{1}{\sqrt{d}}\sum_{n=-s}^{s}{\rm e}^{\frac{2\pi {\rm i}}{d}kn}|n\rangle 
\]
and to define the `momentum' operator 
\[
\begin{array}{l}
P:\mathcal{H}\longrightarrow \mathcal{H},\qquad P=\sqrt{\frac{2\pi }{d}}\sum\limits_{k=-s}^{s}k\, |\tilde k\rangle \langle \tilde k|.
\end{array}
\]
The operators $Q$ and $P$ have the same spectrum, namely,
\[
\mathcal{S}_d=\left\{ -s\sqrt{\frac{2\pi }{d}},\ (-s\!+\!1)\sqrt{\frac{2\pi }{d}},\ ...\ ,\ (s\!-\!1)\sqrt{\frac{2\pi }{d}},\ s\sqrt{\frac{2\pi }{d}} \right\}.
\]
Since
\[
\lim _{d\rightarrow \infty }\sqrt{\frac{2\pi }{d}}=0\qquad {\rm and}\qquad 
 \lim _{d\rightarrow \infty }(\pm s)\sqrt{\frac{2\pi }{d}}=\pm \infty
\]
in the limit $d\rightarrow \infty $, the spectra of $Q$ and $P$ correspond in a certain sense to $(-\infty ,\infty )$.

Each state $|\psi \rangle \!\in \!\mathcal{H}$ can be expanded as
\[
|\psi \rangle \!=\!\sum_{n=-s}^{s}\psi (n)\, |n\rangle \!=\!\sum_{k=-s}^{s}\tilde \psi (k)\, |\tilde k\rangle 
\]
where the functions \
$\psi \!:\!\mathbb{Z}_d\!\longrightarrow \!\mathbb{C}\!:n\mapsto \psi (n)$ \ and \  
$\tilde \psi  \!:\!\mathbb{Z}_d\!\longrightarrow \!\mathbb{C}\!:k\mapsto \tilde \psi (k)$ \ 
satisfying 
\begin{equation} \fl 
\psi (n)\!=\!\langle n|\psi \rangle \!=\!\frac{1}{\sqrt{d}}\sum_{k=-s}^{s}{\rm e}^{\frac{2\pi {\rm i}}{d} kn}\tilde \psi (k)\,,\qquad 
\tilde \psi (k)\!=\!\langle \tilde k|\psi \rangle \!=\!\frac{1}{\sqrt{d}}\sum_{n=-s}^{s}{\rm e}^{-\frac{2\pi {\rm i}}{d}kn}\psi (n)\, .
\end{equation}
are the corresponding `wavefunctions' in the position and momentum representations \cite{cotfas}.
The operators $Q$ and $P$ satisfy the relations 
\[
FQF^+\!=\!P\qquad \qquad FPF^+\!=\!-Q .
\]
The displacement operators  $A\, ,B\!:\!\mathcal{H}\!\longrightarrow \!\mathcal{H}$
\begin{equation} 
 A\!=\!{\rm e}^{-{\rm i}\sqrt{2\pi /d}\, P}\!=\!\!\sum_{k =-s}^s{\rm e}^{-\frac{2\pi {\rm i}}{d}k } |\tilde k\rangle \langle \tilde k|\,,\qquad B\!=\!{\rm e}^{{\rm i}\sqrt{2\pi /d}\, Q}\!=\!\sum_{\ell =-s}^s{\rm e}^{\frac{2\pi {\rm i}}{d}\ell } |\ell\rangle \langle \ell|\, , 
\end{equation}
are single-valued and satisfy the relations
\[
\begin{array}{lll}
A|\ell\rangle =|\ell \!+\!1\rangle \, , & A|\tilde k\rangle \!=\!{\rm e}^{-\frac{2\pi {\rm i}}{d}k}|\tilde k\rangle  \, ,\qquad  & A^d=B^d=\mathbb{I}\, , \\[2mm]
B|\ell\rangle \!=\!{\rm e}^{\frac{2\pi {\rm i}}{d}\ell}|\ell\rangle\, ,  \qquad  & B|\tilde k\rangle =|\widetilde {k\!+\!1}\rangle  \, ,& 
A^\alpha B^\beta \!=\!{\rm e}^{-\frac{2\pi {\rm i}}{d}\alpha \beta }B^\beta A^\alpha\,  .
\end{array}
\]
The general displacements operators \cite{Vourdas2004, ST}
\begin{equation}
D(\alpha,\beta)={\rm e}^{\frac{\pi {\rm i}}{d}\alpha \beta }\, A^\alpha B^\beta \qquad {\rm where}\quad (\alpha,\beta) \in \mathbb{Z}_d \times \mathbb{Z}_d
\end{equation}
define a projective representation of the finite Weyl group. The vectors $\{|\alpha ,\beta \rangle \}_{\alpha ,\beta =-s}^s$, where
\[
|\alpha ,\beta \rangle =D(\alpha ,\beta )\, \frac{\mathfrak{g}_1}{||\mathfrak{g}_1||}=\frac{{\rm e}^{-\frac{\pi {\rm i}}{d}\alpha \beta }}{|| \mathfrak{g}_1||}
\sum_{j=-s}^{s}{\rm e}^{\frac{2\pi {\rm i}}{d}\beta j} \,\,  \mathfrak{g}_1(j\!-\!\alpha )\, |j\rangle 
\]
satisfy the  resolution of identity\cite{cotfas,Z}
\[
\frac{1}{d}\sum _{\alpha ,\beta =-s}^s|\alpha ,\beta \rangle \langle \alpha ,\beta |=\mathbb{I}.
\]
The tight frame $\{|\alpha ,\beta \rangle \}_{\alpha ,\beta =-s}^s$ can be regarded as a finite system of coherent states \cite{G2009} labeled by using the set ${Z}_d \!\times \!\mathbb{Z}_d$, directly related to the {\em finite phase space}
$\mathcal{S}_d\!\times \!\mathcal{S}_d$.

The mathematical objects defined above correspond in the large $d$ limit to those usually considered in the case of the quantum harmonic oscillator. An extensive list concerning this correspondence can be found in Table \ref{corresp}.\\

\begin{table}
\caption{\label{corresp}Correspondence between continuous and $(2s\!+\!1)$-dimensional case.}
\begin{indented}
\lineup
\item[]\begin{tabular}{cc}
\br
${Continuous\ case}$ &   $(2s\!+\!1)-{dimensional\ case}$\\
\mr
$(-\infty ,\infty )$ &  $\mathcal{S}_d$\\[2mm]
$x$ &  $n\sqrt{\frac{2\pi }{d}}$\\[2mm]
$p$ &  $k\sqrt{\frac{2\pi }{d}}$\\[2mm]
$\langle x|x'\rangle =\delta (x-x')$ &  $\langle n|n'\rangle =\delta _{nn'}$\\[2mm]
$|\psi \rangle  =\int dx\, \psi (x)\, |x\rangle $ &  $|\psi \rangle =\sum_{n=-s}^{s}\psi (n)\, |n\rangle $\\[2mm]
$\psi (x)=\langle x|\psi \rangle $ & $\psi (n)=\langle n|\psi \rangle $\\[2mm]
$\hat x=\int dx\, x\, |x\rangle \langle x|$ &  $Q=\sqrt{\frac{2\pi }{d}}\sum\limits_{n=-s}^{s}n\, |n\rangle \langle n|$\\[2mm]
$\hat x \, |x\rangle =x\, |x\rangle $ &  $Q\, |n\rangle =n\sqrt{\frac{2\pi }{d}}\, |n\rangle $\\[2mm]
$\mathcal{F}=\frac{1}{\sqrt{2\pi }}\int \!\!\int dx'\,  dx\, {\rm e}^{{\rm i}x'x}|x'\rangle \langle x|$ &  $F=\frac{1}{\sqrt{d}}\sum_{n',n=-s}^{s}{\rm e}^{\frac{2\pi {\rm i}}{d}n'n}|n'\rangle \langle n|$\\[2mm]
$\mathcal{F}|\psi \rangle =\frac{1}{\sqrt{2\pi }}\int \!\!\int dx'\,  dx\, {\rm e}^{{\rm i}x'x}\psi (x)\, |x'\rangle$    &     $F|\psi \rangle =\frac{1}{\sqrt{d}}\sum_{n',n=-s}^{s}{\rm e}^{\frac{2\pi {\rm i}}{d}n'n}\psi (n)\, |n'\rangle $\\[2mm]
$\mathcal{F}|\psi \rangle =\int dx'\, \mathcal{F}[\psi ](x')\, |x'\rangle $ &   $\mathcal{F}|\psi \rangle =\sum_{n'=-s}^s \mathcal{F}[\psi ](n')\, |n'\rangle  $\\[2mm]
$\mathcal{F}[\psi ](x')=\frac{1}{\sqrt{2\pi }}\int dx\, {\rm e}^{{\rm i}x'x}\psi (x)$ &  $F[\psi ] (n')=\frac{1}{\sqrt{d}}\sum_{n=-s}^{s}{\rm e}^{\frac{2\pi {\rm i}}{d}n'n}\psi (n)$\\[2mm]
$|\tilde p\rangle =\mathcal{F}|p\rangle =\frac{1}{\sqrt{2\pi }}\int  dx\, {\rm e}^{{\rm i}px}|x\rangle $ &  $|\tilde k\rangle =\mathcal{F}|k\rangle =\frac{1}{\sqrt{d }}\sum_{n=-s}^s {\rm e}^{\frac{2\pi {\rm i}}{d}kn}|n\rangle $\\[2mm]
$\langle x|\tilde p\rangle = \frac{1}{\sqrt{2\pi }}{\rm e}^{{\rm i}px}$ & $\langle n|\tilde k\rangle =\frac{1}{\sqrt{d }}{\rm e}^{\frac{2\pi {\rm i}}{d}kn}$\\[2mm]
$\langle \tilde p|\tilde p'\rangle =\delta (p-p')$ &  $\langle \tilde k|\tilde k'\rangle =\delta _{kk'}$\\[2mm]
$|\psi \rangle =\int dp\, \tilde \psi (p)\, |\tilde p\rangle $ & $ |\psi \rangle =\sum_{n=-s}^{s}\tilde \psi (k)\, |\tilde k\rangle $\\[2mm]
$\tilde \psi (p)=\langle \tilde p|\psi \rangle =\frac{1}{\sqrt{2\pi }}\int  dx\, {\rm e}^{-{\rm i}px}\psi (x)$ & $\tilde \psi (k)=\langle \tilde k|\psi \rangle =\frac{1}{\sqrt{d}}\sum_{n=-s}^s {\rm e}^{-\frac{2\pi {\rm i}}{d}kn}\psi (n) $ \\[2mm]
$ \hat p=\int dp\, p\, |\tilde p\rangle \langle \tilde p|$ &  $P=\sqrt{\frac{2\pi }{d}}\sum\limits_{k=-s}^{s}k\, |\tilde k\rangle \langle \tilde k|$\\[2mm]
$\hat p \, |\tilde p\rangle =p\, |\tilde p\rangle $ &  $P\, |\tilde k\rangle =k\sqrt{\frac{2\pi }{d}}\, |\tilde k\rangle $\\[2mm]
$\psi _\kappa :\mathbb{R}\longrightarrow \mathbb{R}$ & $\mathfrak{g}_\kappa :\mathbb{Z}_d\longrightarrow \mathbb{R}$\\[2mm]
$\psi _\kappa (x)=\mathrm{e}^{-\frac{\kappa }{2}x^2}$  &  $ \mathfrak{g}_\kappa  (n)=\sum_{\alpha =-\infty }^\infty \mathrm{e}^{-\frac{\kappa }{2}\left(\sqrt{\frac{2\pi }{d}}\, (\alpha d +n)\right)^2}$\\[2mm]
${\rm e}^{-{\rm i}\alpha \hat p}|x\rangle =|x\!+\!\alpha \rangle  $
&  $A^\alpha |n\rangle =|n \!+\!\alpha \rangle  $\\[2mm]
${\rm e}^{-{\rm i}\alpha \hat p}|\tilde p\rangle ={\rm e}^{-{\rm i}\alpha p}|\tilde p \rangle  $ & $ A^\alpha |\tilde k\rangle \!=\!{\rm e}^{-\frac{2\pi {\rm i}}{d}\alpha k}|\tilde k\rangle $\\[2mm]
${\rm e}^{{\rm i}\beta \hat x}|x\rangle ={\rm e}^{{\rm i}\beta  x}|x\rangle $ &  $B^\beta |n\rangle \!=\!{\rm e}^{\frac{2\pi {\rm i}}{d}\beta n}|n\rangle $\\[2mm]
${\rm e}^{{\rm i}\beta \hat x}|\tilde p\rangle =|\widetilde {p\!+\!\beta}\rangle $  &  $B^\beta |\tilde k\rangle =|\widetilde {k\!+\!\beta}\rangle  $ \\[2mm]
$D(\frac{\alpha \!+\!{\rm i}\beta }{\sqrt{2}})\!=\!{\rm e}^{\frac{\rm i}{2}\alpha \beta }\, {\rm e}^{-{\rm i}\alpha \hat p}\, {\rm e}^{{\rm i}\beta \hat x}$ &  $D(\alpha,\beta)={\rm e}^{\frac{\pi {\rm i}}{d}\alpha \beta }\, A^\alpha B^\beta $ \\[2mm]
$|z\rangle =D(z)\, \frac{\psi _1}{||\psi _1||}$ &  $|\alpha ,\beta \rangle =D(\alpha ,\beta )\, \frac{\mathfrak{g}_1}{||\mathfrak{g}_1||}$ \\[2mm]
$\frac{1}{\pi }\int _\mathbb{C}dz\, |z\rangle \langle z|=\mathbb{I}$  &  $\frac{1}{d}\sum _{\alpha ,\beta =-s}^s|\alpha ,\beta \rangle \langle \alpha ,\beta |=\mathbb{I}$ \\[2mm]
$\mathbb{R}^2$  &  $\mathcal{S}_d\!\times \!\mathcal{S}_d$\\
\br
\end{tabular}
\end{indented}
\end{table}

\noindent {\bf Example}.   The Hamiltonian
\[
H_{\rm free}:\mathcal{H}\longrightarrow \mathcal{H}, \qquad H_{\rm free}=\frac{1}{2}P^2
\]
admits the non-degenerated ground level
\[
\lambda _0=0\qquad {\rm with}\quad |\tilde 0\rangle \quad {\rm a\ corresponding\ eigenvector}  
\]
and the double-degenerated energy levels
\[
\begin{array}{llll}
\lambda _1=\frac{\pi }{d}\, 1^2 & {\rm with} & |\pm \tilde 1\rangle  & {\rm orthogonal\ eigenvectors}\\
\lambda _2=\frac{\pi }{d}\, 2^2 & {\rm with} & |\pm \tilde 2\rangle  & {\rm orthogonal\ eigenvectors}\\
\dots &\dots &\dots &\dots \\
\lambda _s=\frac{\pi }{d}\, s^2 & {\rm with} & |\pm \tilde s\rangle  & {\rm orthogonal\ eigenvectors}.
\end{array}
\]
Therefore,
\[
H_{\rm free}=\frac{\pi }{d}\sum_{n=-s}^sn^2\, |\tilde n\rangle \langle \tilde n| 
\]
and the evolution operator
\[
{\rm e}^{-{\rm i}tH_{\rm free}}=\sum_{n=-s}^s{\rm e}^{-{\rm i}t\frac{\pi }{d}n^2} |\tilde n\rangle \langle \tilde n|
\]
is periodic with period $2d$
\[
{\rm e}^{-{\rm i}(t+2d)H_{\rm free}}  ={\rm e}^{-{\rm i}tH_{\rm free}}.  
\]
For any state $|\psi \rangle \!\in \!\mathcal{H}$, the corresponding time dependent state
\[
\Psi \!:\!\mathbb{Z}_d\times \mathbb{R}\!\longrightarrow \!\mathbb{C}, \qquad 
\Psi (n,t)=\langle n|{\rm e}^{-{\rm i}tH_{\rm free}}|\psi \rangle
\]
is periodic in time
\[
\Psi (n,t+2d)=\Psi (n,t).
\]
Note that in the continuum limit, when $d$ tends to infinity, the periodicity of free evolution practically disappears, and one obtains the result known from continuous-configuration quantum mechanics.

A similar periodicity has been obtained in \cite{Bang2009} for a free wavepacket moving in a discrete quantum phase space. However, in \cite{Bang2009} the discrete-eigenvalues position and momentum operators were defined differently, with the result that the revivals appear for minimum uncertainty states but are only approximate for long time evolution in other cases. 

\bigskip
%
%
%
%%%%%%%%%%%%%%%%%%%%%%%%%%%%
\section{On the commutator $[Q,P]$}\label{Comm}
\bigskip
In this section we present a result similar to Floratos \cite{Floratos}, in a version adapted to our operators $Q$ and $P$, and a numerical 
estimation  of the spectrum of $[Q,P]$.
The matrices of $Q$ and $F$ 
in the basis $\{ |\ell \rangle \}_{\ell \in \mathbb{Z}_d}$ are
\[
\begin{array}{l}
\left( q_{j\ell }\right)_{j,\ell =-s}^s\!=\!\left(\sqrt{\frac{2\pi }{d}}\,j\, \delta _{j\ell }\right)_{j,\ell =-s}^s\qquad 
\left( F_{j\ell }\right)_{j,\ell =-s}^s\!=\!\left( \frac{1}{\sqrt{d}}\,{\rm e}^{\frac{2\pi {\rm i}}{d}j\ell }\right)_{j,\ell =-s}^s
\end{array}
\]
Since $P=FQF^+$ the matrix elements of $P$ in the same basis are
\[
\begin{array}{l}
p_{j\ell }=(F\hat QF^+)_{jk}=\frac{1}{d}\sqrt{\frac{2\pi }{d}}\sum_{k=-s}^sk\,{\rm e}^{\frac{2\pi {\rm i}}{d}(j-\ell )k}.
\end{array} 
\]
If we differentiate with respect to $x$ the identity
\[
\sum_{k=-s}^s{\rm e}^{kx}={\rm e}^{-sx}\, \frac{{\rm e}^{dx}-1}{{\rm e}^{x}-1}
\]
true for $x\neq 0$, then we get the relation
\[
\sum_{k=-s}^sk\, {\rm e}^{kx}=-s\, {\rm e}^{-sx}\, \frac{{\rm e}^{dx}-1}{{\rm e}^{x}-1}+{\rm e}^{-sx}\, \frac{d\, {\rm e}^{dx}({\rm e}^{x}-1)-{\rm e}^{x}({\rm e}^{dx}-1)}{({\rm e}^{x}-1)^2}
\]
which for $x=\frac{2\pi {\rm i}}{d}(j-\ell )$ becomes
\[
\sum_{k=-s}^sk\,{\rm e}^{\frac{2\pi {\rm i}}{d}(j-\ell )k}={\rm e}^{-\frac{2\pi {\rm i}}{d}(j-\ell )s}\, \frac{d}{{\rm e}^{\frac{2\pi {\rm i}}{d}(j-\ell )}-1}.
\]
Since $s=\frac{d-1}{2}$, the last relation can be written as
\[
\sum_{k=-s}^sk\,{\rm e}^{\frac{2\pi {\rm i}}{d}(j-\ell )k}=(-1)^{j-\ell }\, 
\frac{d}{{\rm e}^{\frac{\pi {\rm i}}{d}(j-\ell )}-{\rm e}^{-\frac{\pi {\rm i}}{d}(j-\ell )}}
\]
and we have
\[
p_{j\ell }=\left\{ 
\begin{array}{cll}
0 & {\rm if} & j\!=\!\ell \\
-\frac{\rm i}{2}\sqrt{\frac{2\pi }{d}}\, \frac{(-1)^{j-\ell }}{\sin \frac{\pi }{d}(j-\ell )}& {\rm if} & j\!\neq \!\ell .
\end{array}\right. 
\]
The matrix elements of the commutator $[Q,P]$ are
\begin{equation}\label{qppq}
[Q,P]_{j\ell}=\left\{ 
\begin{array}{cll}
0 & {\rm if} & j\!=\!\ell \\
-{\rm i}\, \frac{\frac{\pi }{d}(j-\ell )\, (-1)^{j-\ell }}{\sin \frac{\pi }{d}(j-\ell )}& {\rm if} & j\!\neq \!\ell 
\end{array}\right. 
\end{equation}
and for large $d$, they  can be approximated as follows \cite{Floratos}
\[
[Q,P]_{j\ell }\approx {\rm i}\,(-1)^{j-\ell }(\delta _{j\ell}-1).
\]
The matrix 
\[
\begin{array}{l}
\left({\rm i}\, (-1)^{j-\ell }(\delta _{j\ell}-1)\right)_{j,\ell =-s}^s
\end{array}
\]
has the eigenvectors
\[ \fl 
\begin{array}{l}
 \left( \frac{1}{\sqrt{d}}(-1)^j  {\rm e}^{\frac{2\pi {\rm i}}{d}j k}\right)_{j=-s}^s\  
{\rm with\  eigenvalue}\ \left\{ \!\!
\begin{array}{rll}
{\rm i} & {\rm for} & k\!\in \!\{ -s,...,-1,1,...,s\}\\[3mm]
(1\!-\!d){\rm i} & {\rm for} & k\!=\!0.
\end{array} \right.
\end{array}
\]
This means that $d\!-\!1$ of the eigenvalues of $[Q,P]$ are equal to ${\rm i}$
for large $d$. We can consider that, in a certain sense, 
\[
[Q,P]\approx {\rm i}.
\]
A numerical estimation of the eigenvalues of the commutator $[Q,P]$  in the case $d=15$  can be seen in Table \ref{commutator-eigenval}.  Already for this relative small $d$ value a significant number of eigenvalues tend to i.

\begin{table}
\caption{\label{commutator-eigenval}The eigenvalues $\eta _n$ of the commutator $[Q,P]$ in the case $d=15$.}
\begin{indented}
\lineup
\item[]\begin{tabular}{rrrrrr}
\br
$k$ & $\eta _k\qquad \quad $  & $k$ & $\eta _k\qquad \quad $  & $k$ & $\eta _k\qquad \quad $  \\
\mr
0 &  -27.276466375122 \,{\rm i}& 5 & 0.999998706977 \,{\rm i}& 10 & 1.000016906603 \,{\rm i}\\
1 &  -4.322222514423 \,{\rm i}& 6 & 0.999999996717 \,{\rm i}& 11 & 1.001534631543 \,{\rm i}\\
2 &  0.649632619978 \,{\rm i}& 7 & 0.999999999998 \,{\rm i}& 12 & 1.067898771074 \,{\rm i}\\
3 & 0.988901431861 \,{\rm i}& 8 & 1.000000000091 \,{\rm i}& 13 & 2.560890405316 \,{\rm i}\\
4 & 0.999822475466 \,{\rm i}& 9 & 1.000000076444 \,{\rm i}& 14 & 18.32999286747 \,{\rm i}\\
\br
\end{tabular}
\end{indented}
\end{table}

\bigskip
%
%%%%%%%%%%%%%%%%%%%%%%%%%%%%%%%
\section{Minimum uncertainty states}\label{Mus}
\bigskip
Let $\kappa \!\in \!(0,\infty )$, \  $s\!\in \!\{ 1,2,3,...\}$ and let $d\!=\!2s\!+\!1$. The function
\[
G_\kappa :\mathbb{R}\longrightarrow \mathbb{R},\qquad G_\kappa (x)=\sum_{\alpha =-\infty }^\infty \mathrm{e}^{-\frac{\kappa }{2}\left(\sqrt{\frac{2\pi }{d}}\, (\alpha d +x)\right)^2}=\sum_{\alpha =-\infty }^\infty \mathrm{e}^{-\frac{\kappa \pi }{d} \left(\alpha d +x\right)^2}
\]
obtained by starting from the Gaussian function 
\[
g_\kappa :\mathbb{R}\longrightarrow \mathbb{R},\qquad g_\kappa (x)=\mathrm{e}^{-\frac{\kappa }{2}x^2}\\[-1mm]
\]
is a periodic function with period $d$. The finite Gaussian
\[
\mathfrak{g}_\kappa  :\mathbb{Z}_d\longrightarrow \mathbb{R},\qquad \mathfrak{g}_\kappa  (n)=\sum_{\alpha =-\infty }^\infty \mathrm{e}^{-\frac{\kappa \pi }{d} (\alpha d +n)^2}\\[-1mm]
\]
which can be regarded as a finite version of $g_\kappa $, satisfies the relation
\[
\mathfrak{g}_\kappa (-n)=\mathfrak{g}_\kappa (n),\qquad {\rm for\ any}\quad n\in \mathbb{Z}_d.
\]
{\bf Theorem 1} \cite{Ruzzi}. {\it We have}
\begin{equation}\label{fourierg}
\begin{array}{l}
F[\mathfrak{g}_\kappa ]=\frac{1}{\sqrt{\kappa }}\, \mathfrak{g}_{\frac{1}{\kappa }}\qquad for\ any\quad \kappa \!\in \!(0,\infty ).\\[3mm]
\end{array}
\end{equation}
{\bf Proof.}
 The function $G_\kappa (x)$ admits the Fourier expansion\\[-3mm]
\[
G_\kappa (x)=\sum_{\ell =-\infty }^\infty a_\ell \,  \mathrm{e}^{\frac{2\pi \mathrm{i}}{d}\ell x}\\[-1mm]
\]
with\\[-3mm]
\[ \fl
a_\ell =\frac{1}{d}\int _0^d\mathrm{e}^{-\frac{2\pi \mathrm{i}}{d}\ell x}\sum_{\alpha =-\infty }^\infty \mathrm{e}^{-\frac{\kappa }{2}\left(\sqrt{\frac{2\pi }{d}}\, (\alpha d +x)\right)^2}\, dx=\frac{1}{d}\sum_{\alpha =-\infty }^\infty\int _0^d\mathrm{e}^{-\frac{2\pi \mathrm{i}}{d}\ell x} \mathrm{e}^{-\frac{\kappa }{2}\left(\sqrt{\frac{2\pi }{d}}\, (\alpha d +x)\right)^2}\, dx
\\[-1mm]
\]
By denoting $t\!=\!\sqrt{\frac{2\pi }{d}}\, (\alpha d \!+\!x)$ and using the relation
\[
\int_{-\infty }^{\infty}\mathrm{e}^{\mathrm{i}\xi t}\,  \mathrm{e}^{-a t^2}\, dt=\sqrt{\frac{\pi }{a }}\,\,  \mathrm{e}^{-\frac{\xi ^2}{4a }}
\]
we get \cite{Mehta} \\[-3mm]
\[
\begin{array}{rl}
a_\ell & =\frac{1}{\sqrt{2\pi d}}\sum_{\alpha =-\infty }^\infty \int_{\alpha \sqrt{2\pi d}}^{(\alpha +1)\sqrt{2\pi d}}\mathrm{e}^{-\frac{2\pi \mathrm{i}}{d}\ell \left(t\sqrt{\frac{d}{2\pi }}-\alpha d\right)}
\mathrm{e}^{-\frac{\kappa }{2}t^2}\, dt\\[4mm]
 & =\frac{1}{\sqrt{2\pi d}}\sum_{\alpha =-\infty }^\infty \int_{\alpha \sqrt{2\pi d}}^{(\alpha +1)\sqrt{2\pi d}}\mathrm{e}^{-\mathrm{i}\ell  t\sqrt{\frac{2\pi }{d}}}
\mathrm{e}^{-\frac{\kappa }{2}t^2}\, dt\\[4mm]
& =\frac{1}{\sqrt{2\pi d}}\int_{-\infty }^{\infty}\mathrm{e}^{-\mathrm{i}\ell  t\sqrt{\frac{2\pi }{d}}} \mathrm{e}^{-\frac{\kappa }{2}t^2}\,  dt=\frac{1}{\sqrt{\kappa d}}\, \mathrm{e}^{-\frac{\pi}{\kappa d}\ell ^2}\\[-1mm]
\end{array}
\]
whence\\[-3mm]
\[
G_\kappa (x)=\frac{1}{\sqrt{\kappa d}}\,\sum_{\ell =-\infty }^\infty   \mathrm{e}^{\frac{2\pi \mathrm{i}}{d}\ell x}\, \mathrm{e}^{-\frac{\pi}{\kappa d}\ell ^2}.\\[-1mm]
\]
Particularly, we have\\[-3mm]
\[
\begin{array}{rl}
\mathfrak{g}_\kappa (j) & =G_\kappa (j)=\frac{1}{\sqrt{\kappa d}}\,\sum_{\ell =-\infty }^\infty   \mathrm{e}^{\frac{2\pi \mathrm{i}}{d}j\ell }\, \mathrm{e}^{-\frac{\pi}{\kappa d}\ell ^2}\\[3mm]
 & =\frac{1}{\sqrt{\kappa d}}\sum_{n=-s}^{s}\sum_{\alpha =-\infty }^\infty  \mathrm{e}^{\frac{2\pi \mathrm{i}}{d}j(\alpha d+n) }\, \mathrm{e}^{-\frac{\pi }{\kappa d}(\alpha d+n) ^2}\, \\[2mm]
& =\frac{1}{\sqrt{\kappa }}\frac{1}{\sqrt{d}}\sum_{n=-s}^{s}\mathrm{e}^{\frac{2\pi \mathrm{i}}{d}jn }\sum_{\alpha =-\infty }^\infty  \, \mathrm{e}^{-\frac{\pi }{\kappa d}(\alpha d+n) ^2}
\end{array}
\]
whence\\[-3mm] 
\[
\frac{1}{\sqrt{d}}\sum_{n=-s}^{s}\mathrm{e}^{\frac{2\pi \mathrm{i}}{d}jn }\sum_{\alpha =-\infty }^\infty  \, \mathrm{e}^{-\frac{\pi }{\kappa d}(\alpha d+n) ^2}=\sqrt{\kappa }\, \sum_{\alpha =-\infty }^\infty  \, \mathrm{e}^{-\frac{\kappa \pi }{d}(\alpha d+j) ^2}\qquad \opensquare
\]

Since $\mathfrak{g}_\kappa (-n)=\mathfrak{g}_\kappa (n)$ and $\mathfrak{g}_{\frac{1}{\kappa }} (-n)=\mathfrak{g}_{\frac{1}{\kappa }} (n)$ we have
\[
\sum_{n=-s}^sn(\mathfrak{g}_\kappa (n))^2=\sum_{n=-s}^sn(\mathfrak{g}_{\frac{1}{\kappa }} (n))^2=0.
\]
Therefore, the square of the dispersion of $Q$ in the state described by the finite Gaussian 
\[
\mathfrak{g}_\kappa  :\mathbb{Z}_d\longrightarrow \mathbb{R},\qquad \mathfrak{g}_\kappa  (n)=\sum_{\alpha =-\infty }^\infty \mathrm{e}^{-\frac{\kappa \pi }{d} (\alpha d +n)^2}\\[-1mm]
\]
is
\[
(\Delta Q)^2=\langle Q^2\rangle -\langle Q\rangle ^2=\frac{2\pi }{d}\frac{\sum_{n=-s}^sn^2(\mathfrak{g}_\kappa (n))^2}{\sum_{n=-s}^s(\mathfrak{g}_\kappa (n))^2}
\]
and, in view of the relation $P\!=\!FQF^+$, the square of the dispersion of $P$ is
\[
(\Delta P)^2=\langle P^2\rangle -\langle P\rangle ^2=\frac{2\pi }{d}\frac{\sum_{n=-s}^sn^2(\mathfrak{g}_{\frac{1}{\kappa }}(n))^2}{\sum_{n=-s}^s(\mathfrak{g}_{\frac{1}{\kappa }}(n))^2}.
\]
We have
\[
\Delta Q\, \Delta P=\frac{2\pi }{d}\sqrt{\frac{\sum_{n=-s}^sn^2(\mathfrak{g}_\kappa (n))^2}{\sum_{n=-s}^s(\mathfrak{g}_\kappa (n))^2}}\,  \sqrt{\frac{\sum_{n=-s}^sn^2(\mathfrak{g}_{\frac{1}{\kappa }}(n))^2}{\sum_{n=-s}^s(\mathfrak{g}_{\frac{1}{\kappa }}(n))^2}}.
\]

\begin{table}
\caption{\label{uncertainty}The state $\mathfrak{g}_1$ is a quasi-minimum uncertainty state.}
\begin{indented}
\lineup
\item[]\begin{tabular}{rlll}
\br
$d$ & $\Delta Q\, \Delta P$  & $\frac{1}{2}\, |\langle [Q,P]\rangle |$  & $\Delta Q\, \Delta P- \frac{1}{2}\, |\langle [Q,P]\rangle |$ \\
\mr
3 & 0.44259776311852& 0.44259776311852 &$ 0.0 \cdot 10^{-75}$ \\
5 & 0.49709993841560 & 0.49620649757954 &   0.000893440\\
7 & 0.49985914364743 & 0.49985140492777 & $7.738719663 \cdot 10^{-6}$\\
9 &   0.49999327972581 & 0.49999098992968 & $2.289796128 \cdot 10^{-6}$\\
11 & 0.49999968416091 & 0.49999965440967 & $2.975123667 \cdot 10^{-8}$\\
13 & 0.49999998532738 & 0.49999998026367 & $5.063715121 \cdot 10^{-9}$\\
15 & 0.49999999932443 & 0.49999999924381 & $8.061781262 \cdot 10^{-11}$\\
\br
\end{tabular}
\end{indented}
\end{table}

\noindent As concerns the expectation value of $[Q,P]$ , by using the relation (\ref{qppq}), we obtain
\[ \fl
\langle [Q,P]\rangle \!=\!\frac{\langle \mathfrak{g}_\kappa |[Q,P]|\mathfrak{g}_\kappa \rangle }{\langle \mathfrak{g}_\kappa |\mathfrak{g}_\kappa \rangle }\!=\!\frac{2{\rm i}}{\sum_{n=-s}^s(\mathfrak{g}_\kappa (n))^2}\!\left( \sum_{j=-s+1}^s\sum_{\ell =-s}^{j-1}\! (-1)^{j-\ell }\, \frac{\frac{\pi }{d}(j\!-\!\ell )}{\sin \frac{\pi }{d}(j-\ell)}\, \mathfrak{g}_\kappa (j)\, \mathfrak{g}_\kappa (\ell ) \right). 
\]
The well-known uncertainty relation originating from Schwarz inequality
\begin{equation}\label{uncert1}
\Delta Q\, \Delta P\geq \frac{1}{2}\, |\langle [Q,P]\rangle |
\end{equation}
is satisfied, and for $\kappa =1$  the difference 
\begin{equation}\label{uncert2}
\Delta Q\, \Delta P- \frac{1}{2}\, |\langle [Q,P]\rangle |\approx 0
\end{equation}
except a few  small values of $d$. Numerical results concerning the case $\kappa =1$ are presented in table \ref{uncertainty}.

Note that for different definitions of the position and momentum operators  in a discrete quantum phase space \cite{Bang2009}, the minimum uncertainty of these operators is dependent on the discretization step (the exact formula is known as the generalized uncertainty principle) and approaches the result for the continuum case only for an infinitely fine discretization; the generalized uncertainty principle can be obtained from a quantum mechanical model with discrete eigenvalues for the coordinate operator \cite{Bang2006}. On the contrary, in our case the uncertainty is approximately minimum for quite small $d$ values. Moreover, in \cite{massar} the uncertainty relation was shown to reach its minimum value only in particular cases, but approximate expansions of the unitary position and momentum operators on particular states were used. 
\bigskip
%
%%%%%%%%%%%%%%%%%%%%%%%%%%%%%%%
\section{Finite-dimensional quantum system of oscillator type}\label{Sot}
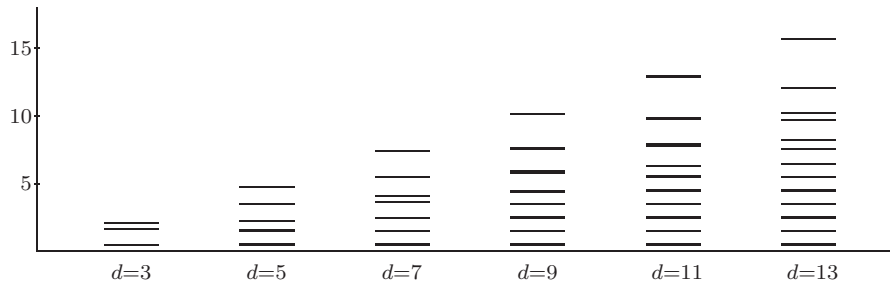
\begin{figure}[h]
\begin{picture}(70,120) (-45,-5)
\setlength{\unitlength}{1.8mm}
\put(5,0){\line(1,0){63}}
\put(4.8,5){\line(1,0){0.4}}
\put(4.8,10){\line(1,0){0.4}}
\put(4.8,15){\line(1,0){0.4}}
\put(3.8,4.5){$\scriptstyle{5}$}
\put(3,9.5){$\scriptstyle{10}$}
\put(3,14.5){$\scriptstyle{15}$}
\put(5,0){\line(0,1){18}}
\put(10.5,-2){$\scriptstyle{d=3}$}
\put(10,2.0943951023931954923){\line(1,0){4}} 
\put(10,1.6517973392746703630){\line(1,0){4}} 
\put(10,0.44259776311852512929){\line(1,0){4}}
\put(20.5,-2){$\scriptstyle{d=5}$}
\put(20,4.745031651763185909){\line(1,0){4}} 
\put(20,3.512928870280915013){\line(1,0){4}} 
\put(20,2.273277799898969258){\line(1,0){4}}
\put(20,1.538153655416400568){\line(1,0){4}} 
\put(20,0.4969786369997022051){\line(1,0){4}}
\put(30.5,-2){$\scriptstyle{d=7}$}
\put(30,7.43385759445478451){\line(1,0){4}}
\put(30,5.50140576717735412){\line(1,0){4}} 
\put(30,4.09277086004846592){\line(1,0){4}}
\put(30,3.62995143640368874){\line(1,0){4}} 
\put(30,2.47233783699377457){\line(1,0){4}} 
\put(30,1.502561583500699708){\line(1,0){4}}
\put(30,0.499856150139578337){\line(1,0){4}}
\put(40.5,-2){$\scriptstyle{d=9}$}
\put(40,10.1567063512714060){\line(1,0){4}} 
\put(40,7.60184907174441951){\line(1,0){4}} 
\put(40,5.92973728896402282){\line(1,0){4}}
\put(40,5.77295679998478618){\line(1,0){4}} 
\put(40,4.41464563337680779){\line(1,0){4}} 
\put(40,3.51412161417356547){\line(1,0){4}}
\put(40,2.49772584010989943){\line(1,0){4}} 
\put(40,1.50016625850219728){\line(1,0){4}} 
\put(40,0.499993189736805367){\line(1,0){4}}
\put(50.4,-2){$\scriptstyle{d=11}$}
\put(50,12.9088130508553718){\line(1,0){4}}
\put(50,9.8025414079905865){\line(1,0){4}} 
\put(50,7.9646966778296552){\line(1,0){4}} 
\put(50,7.7995168891272831){\line(1,0){4}} 
\put(50,6.3246269976446565){\line(1,0){4}} 
\put(50,5.54102579221970530){\line(1,0){4}}
\put(50,4.48940449755755592){\line(1,0){4}} 
\put(50,3.50138128059791320){\line(1,0){4}} 
\put(50,2.49983706209132176){\line(1,0){4}}
\put(50,1.50000973439528691){\line(1,0){4}} 
\put(50,0.499999681486528590){\line(1,0){4}}
\put(60.4,-2){$\scriptstyle{d=13}$}
\put(60,15.6858062111396956){\line(1,0){4}}
\put(60,12.0888294874935604){\line(1,0){4}} 
\put(60,10.2024626522878073){\line(1,0){4}}
\put(60,9.7134880733084487){\line(1,0){4}} 
\put(60,8.2116879367741713){\line(1,0){4}} 
\put(60,7.5884610505873736){\line(1,0){4}}
\put(60,6.4693456592281102){\line(1,0){4}} 
\put(60,5.5054526489310496){\line(1,0){4}} 
\put(60,4.4989567577622088){\line(1,0){4}}
\put(60,3.50011404063347207){\line(1,0){4}} 
\put(60,2.49998925045441074){\line(1,0){4}} 
\put(60,1.50000054667770710){\line(1,0){4}}
\put(60,0.49999998523619522){\line(1,0){4}}
\end{picture}
\caption{The energy levels of $H\!=\!\frac{1}{2}(P^2\!+\!Q^2)$. }\label{levels}
\end{figure}

\bigskip
The Hamiltonian $H\!=\!\frac{1}{2}(P^2\!+\!Q^2)$ is of harmonic oscillator type, but a certain similitude between the behaviour of our quantum system with finite-dimensional Hilbert space and the standard harmonic oscillator exists only for a large enough dimension $d$. Particularly, the tendency to have equidistant energy levels becomes evident only for $d$ large enough (see Table \ref{eigenvalues} and Figure \ref{levels}).

The finite Gaussian  $\mathfrak{g}_1$ is a quasi-eigenstate of the oscillator type Hamiltonian 
\[
H=\frac{1}{2}(P^2+Q^2).
\]

\begin{table}
\caption{\label{eigenvalues}The eigenvalues of  $H\!=\!\frac{1}{2}(P^2\!+\!Q^2)$.}
\begin{indented}
\lineup
\item[]\begin{tabular}{rrrrrr}
\br
$d=3\ \ \ $  &  $d=5\ \ \ $ & $d=7\ \ \ $  & $d=9\ \ \ $  & $d=11\ \ $&  $d=13\ \ $\\
\mr
 $-\qquad$& $-\qquad$&$-\qquad$ &$-\qquad$ & $-\qquad$& 15.685806\\
 $-\qquad$&$-\qquad$ & $-\qquad$&$-\qquad$ & $-\qquad$& 12.088829\\
$-\qquad$ &$-\qquad$ &$-\qquad$ &$-\qquad$ & 12.908813& 10.202462\\
$-\qquad$ &$-\qquad$ &$-\qquad$ & $-\qquad$& 9.802541& 9.713488\\
$-\qquad$ & $-\qquad$& $-\qquad$& 10.156706& 7.964696& 8.211687\\
 $-\qquad$&$-\qquad$ & $-\qquad$& 7.601849& 7.799516& 7.588461\\
$-\qquad$ &$-\qquad$ & 7.433857& 5.929737& 6.324626& 6.469345\\
 $-\qquad$&$-\qquad$ & 5.501405&5.772956 & 5.541025&5.505452 \\
$-\qquad$& 4.745031& 4.092770& 4.414645&4.489404 &4.498956 \\
$-\qquad$ & 3.512928& 3.629951&3.514121 &3.501381 & 3.500114\\
 2.094395&2.273277 &2.472337 & 2.497725& 2.499837&2.499989 \\
1.651797 &1.538153 &1.502561 &1.500166 & 1.500009& 1.500000\\
0.442597 &0.496978 & 0.499856& 0.499993& 0.499999& 0.499999\\
\br
\end{tabular}
\end{indented}
\end{table}

\begin{table}[t]
\caption{\label{eigenstate}The state $\mathfrak{g}_1$ is a quasi-eigenstate of $H\!=\!\frac{1}{2}(P^2\!+\!Q^2)$.}
\begin{indented}
\lineup
\item[]\begin{tabular}{cccccc}
\br
 &  $d=3$ & $d=5$  & $d=7$  & $d=9$&  $d=11$\\
\mr
$\lambda $ &0.442598 &0.489794 &0.498096 &0.499638 &0.49993\\
$(H\mathfrak{g}_1\!-\!\lambda \mathfrak{g}_1)(\pm1)$& $2.2\!\cdot \!10^{-16}$&$1.2\!\cdot \! 10^{-2}$ &$2.8\!\cdot \!10^{-3}$&$5.8\!\cdot \!10^{-4}$ & $1.1\!\cdot \!10^{-4}$  \\
$(H\mathfrak{g}_1\!-\!\lambda \mathfrak{g}_1)(\pm 2)$&- &$-1.2\!\cdot \!10^{-2}$ & $-8.5\!\cdot \!10^{-4}$&$-1.5\!\cdot \!10^{-4}$ & $-3.1\!\cdot \!10^{-5}$  \\
$(H\mathfrak{g}_1\!-\!\lambda \mathfrak{g}_1)(\pm 3)$&- & -&$-2.0\!\cdot \!10^{-3}$ &$1.3\!\cdot \!10^{-4}$ & $4.3\!\cdot \!10^{-5}$  \\
$(H\mathfrak{g}_1\!-\!\lambda \mathfrak{g}_1)(\pm 4)$& -& -&- &$-5.5\!\cdot \!10^{-4}$ & $-2.9\!\cdot \!10^{-5}$ \\
$(H\mathfrak{g}_1\!-\!\lambda \mathfrak{g}_1)(\pm 5)$& -& -&-&- &$-1.0\!\cdot \!10^{-4}$ \\
\br
\end{tabular}
\end{indented}
\end{table}

We have (see table \ref{eigenstate})
\[
H\mathfrak{g}_1\approx \lambda \mathfrak{g}_1 \qquad \qquad {\rm for}\qquad \qquad \lambda =\frac{H\mathfrak{g}_1\, (0)}{\mathfrak{g}_1(0)}.
\]
This relation can be regarded as an approximative finite version of the relation
\[
\frac{1}{2}\left( -\frac{d^2}{dx^2}+x^2 \right){\rm e}^{-\frac{1}{2}x^2}=\frac{1}{2}\, {\rm e}^{-\frac{1}{2}x^2}.
\]
\newpage
%%%%%%%%%%%%%%%%%%%%%%%%%%%%%%%
\section{On the occurrence of revivals}\label{Or}
{\bf Theorem 2}. {\it If $H$ has $k\geq 2$ commensurate energy levels then there exist revivals.}\\[3mm]
{\bf Proof.}  
If the energy levels $\varepsilon _1$, $\varepsilon _2$, ..., $\varepsilon _k$ are comensurate then
$\varepsilon _2/\varepsilon _1$, \ $\varepsilon _3/\varepsilon _1$, ... , $\varepsilon _k/\varepsilon _1$ are rational numbers and can be represented as fractions.
If $m$ is the least common multiple of the denominators of these 
fractions then there exist the integers $\ell _2$, $\ell _3$, ... , $\ell _k$ such that  
\[
\frac{\varepsilon _j}{\varepsilon _1}=\frac{\ell _j}{m}\in \mathbb{Q} \qquad {\rm for\ any}\ \ j\in \{ 2,3,...,k\}.
\]
If  $|\psi _1\rangle$, $|\psi _2\rangle$, ..., $|\psi _k\rangle$ are the eigenstates corresponding to $\varepsilon _1$, $\varepsilon _2$, ..., $\varepsilon _k$, that is, 
\[
H|\psi _j\rangle =\varepsilon _j\, |\psi _j\rangle \qquad {\rm for\ any}\ \ j\in \{ 1,2,,...,k\}
\]
then the time dependent state 
\[
\Psi \!:\!\mathbb{Z}_d\times \mathbb{R}\!\longrightarrow \!\mathbb{C}, \qquad 
\Psi (n,t)=\langle n|{\rm e}^{-{\rm i}tH}|\psi \rangle
\]
corresponding to an arbitrary state of the form (certain coefficients may be 0)
\[
|\psi \rangle =\alpha  _1|\psi _1\rangle +\alpha _2 |\psi _2\rangle +...+\alpha  _k|\psi _k\rangle 
\]
is periodic with the period $\frac{2m\pi }{\varepsilon _1}$. Indeed, 
\[
\begin{array}{rl}
{\rm e}^{-{\rm i}tH}|\psi \rangle & =\alpha _1{\rm e}^{-{\rm i}t\varepsilon _1} |\psi _1\rangle +\alpha _2 {\rm e}^{-{\rm i}t\varepsilon _2} |\psi _2\rangle +...+\alpha _k {\rm e}^{-{\rm i}t\varepsilon _k} |\psi _k\rangle  \\[2mm]
 & =\alpha _1{\rm e}^{-{\rm i}t\varepsilon _1} |\psi _1\rangle +\alpha _2 {\rm e}^{-{\rm i}t\frac{\ell _2}{m}\varepsilon _1} |\psi _2\rangle +...+\alpha _k {\rm e}^{-{\rm i}t\frac{\ell _k}{m}\varepsilon _1} |\psi _k\rangle  
\end{array}
\]
and we have
\[
{\rm e}^{-{\rm i}(t+\frac{2m\pi }{\varepsilon _1})H}|\psi \rangle ={\rm e}^{-{\rm i}tH}|\psi \rangle 
\]
whence
\[
\Psi \left(n,t+\frac{2m\pi }{\varepsilon _1}\right)=\Psi (n,t).\qquad \opensquare
\]
{\bf Theorem 3}. {\it If $H$ has $k\geq 3$ equidistant energy levels $\varepsilon _1$, $\varepsilon _2$, ..., $\varepsilon _k$, that is , if  
\[
\varepsilon _2-\varepsilon _1=\varepsilon _3-\varepsilon _2=...=\varepsilon _k-\varepsilon _{k-1}
\]
then there exist revivals.}\\[3mm]
{\bf Proof.}  If  $|\psi _1\rangle$, $|\psi _2\rangle$, ..., $|\psi _k\rangle$ are corresponding eigenstates
\[
H|\psi _j\rangle =\varepsilon _j\, |\psi _j\rangle \qquad {\rm for\ any}\ \ j\in \{ 1,2,,...,k\}
\]
then the time dependent state 
\[
\Psi \!:\!\mathbb{Z}_d\times \mathbb{R}\!\longrightarrow \!\mathbb{C}, \qquad 
\Psi (n,t)=\langle n|{\rm e}^{-{\rm i}tH}|\psi \rangle
\]
corresponding to an arbitrary state of the form  (certain coefficients may be 0)
\[
|\psi \rangle =\alpha  _1|\psi _1\rangle +\alpha _2 |\psi _2\rangle +...+\alpha  _k|\psi _k\rangle 
\]
is periodic with the period $\frac{2\pi }{\varepsilon _2-\varepsilon _1}$. Indeed, 
\[
\begin{array}{rl}
{\rm e}^{-{\rm i}tH}|\psi \rangle & \!\!\!=\alpha _1{\rm e}^{-{\rm i}t\varepsilon _1} |\psi _1\rangle +\alpha _2 {\rm e}^{-{\rm i}t\varepsilon _2} |\psi _2\rangle +...+\alpha _k {\rm e}^{-{\rm i}t\varepsilon _k} |\psi _k\rangle  \\[2mm]
 & \!\!\!={\rm e}^{-{\rm i}t\varepsilon _1}\left(\alpha _1 |\psi _1\rangle +\alpha _2 {\rm e}^{-{\rm i}t(\varepsilon _2-\varepsilon _1)} |\psi _2\rangle +...+\alpha _k {\rm e}^{-{\rm i}t(\varepsilon _k-\varepsilon _1)} |\psi _k\rangle  \right)\\[2mm]
 & \!\!\!={\rm e}^{-{\rm i}t\varepsilon _1}\left(\alpha _1 |\psi _1\rangle \!+\!\alpha _2 {\rm e}^{-{\rm i}t(\varepsilon _2-\varepsilon _1)} |\psi _2\rangle \!+...+\!\alpha _k {\rm e}^{-{\rm i}t(k-1)(\varepsilon _2-\varepsilon _1)} |\psi _k\rangle  \right)
\end{array}
\]
and, up to a phase factor, we have 
\[
{\rm e}^{-{\rm i}(t+\frac{2\pi }{\varepsilon _2-\varepsilon _1})H}|\psi \rangle ={\rm e}^{-{\rm i}tH}|\psi \rangle 
\]
whence
\[
\begin{array}{l}
\Psi \left(n,t+\frac{2\pi }{\varepsilon _2-\varepsilon _1}\right)=\Psi (n,t).
\end{array}
\]
The number $\frac{2\pi }{\varepsilon _2-\varepsilon _1}$ is a period for any coefficients $\alpha _1$, $\alpha _2$, ... , $\alpha _k$, but generally, it is not a ``fundamental'' period. For example, in the particular case
\[
|\psi \rangle =\alpha  _1|\psi _1\rangle +\alpha _3 |\psi _3\rangle 
\]
there exists a smaller period, namely,  $\frac{\pi }{\varepsilon _2-\varepsilon _1}.\qquad \opensquare $

In the case of our finite-dimensional oscillator, we have equidistant levels and hence revivals only for $d$ large enough. Once again, the discrete-continuum transition recovers the standard results for the quantum harmonic oscillator. But, more importantly, the proposition above relates the revivals with the condition of equidistant energy levels. From a physical point of view this relation is extremely important: unlike for free evolution, where the revivals’ period was determined by $d$ but the energy spectrum had no equidistant levels and thus the revivals disappeared in the continuous limit, for the harmonic oscillator case $d$ does not explicitly enter the expression of the revivals’ period. This period does not disappear in the large $d$ limit; on the contrary, a large $d$ guarantees equidistant energy levels, which determine a sort of physical feedback necessary for revivals.

%
%%%%%%%%%%%%%%%%%%%%%%%%%%%%%%%%%%
\section{Discrete Wigner function} \label{DWF}
Let $\kappa \!\in \!(0,\infty )$, \  $s\!\in \!\{ 1,2,3,...\}$ and let $d\!=\!2s\!+\!1$. The periodic function $G_\kappa ^+ :\mathbb{R}\longrightarrow \mathbb{R}$
\[ 
G_\kappa ^+ (x)=\sum_{\alpha =-\infty }^\infty \mathrm{e}^{-\frac{\kappa }{2}\left(\sqrt{\frac{2\pi }{d}}\, \left(\left(\alpha + \frac{1}{2}\right)d +x\right)\right)^2}=\sum_{\alpha =-\infty }^\infty \mathrm{e}^{-\frac{\kappa \pi }{d} \left(\left(\alpha + \frac{1}{2}\right)d +x\right)^2}
\]
with period $d$ allows us to define the function $\mathfrak{g}_\kappa ^+  :\mathbb{Z}_d\longrightarrow \mathbb{R}$
\[
\mathfrak{g}_\kappa ^+  (n)=\sum_{\alpha =-\infty }^\infty \mathrm{e}^{-\frac{\kappa \pi }{d} \left(\left(\alpha + \frac{1}{2}\right)d +n\right)^2}=\sum_{\alpha =-\infty }^\infty \mathrm{e}^{-\frac{\kappa \pi }{d} \left(\left(\alpha - \frac{1}{2}\right)d +n\right)^2}.
\]
Since
\[
\begin{array}{l}
\mathfrak{g}_\kappa ^+  (n)=G_\kappa  \left(n\!+\!s\!+\!\frac{1}{2}\right)\qquad {\rm and}\qquad \mathfrak{g}_\kappa  (n)=G_\kappa  \left(n\right)
\end{array}
\]
the function $\mathfrak{g}_\kappa ^+$ is a kind of translated finite Gaussian 
 (see Figure \ref{Wigner}). By direct computation one can prove the relations
\[
 \mathfrak{g}_\kappa ^+ (-n)=\mathfrak{g}_\kappa ^+ (n),\qquad \qquad \mathfrak{g}_\kappa (2n)=\mathfrak{g}_{4\kappa }(n)+\mathfrak{g}_{4\kappa }^+(n)
\]
and
\[
\sum_{\alpha =-\infty }^\infty (-1)^\alpha \mathrm{e}^{-\frac{\kappa \pi }{d} (\alpha d +2n)^2}=\mathfrak{g}_{4\kappa }(n)-\mathfrak{g}_{4\kappa }^+(n).
\]
 {\bf Lemma 1}. {\it The finite Fourier transform of $\mathfrak{g}_{2\kappa }^+ $ satisfies the relation}
\begin{equation}\label{fouriergg}
\mathcal{F}[\mathfrak{g}_{2\kappa }^+ ] (2m)=\frac{1}{\sqrt{2\kappa }}\left( \mathfrak{g}_{\frac{2}{\kappa }} (m)-\mathfrak{g}_{\frac{2}{\kappa }}^+ (m)\right).
\end{equation}
{\bf Proof}. The periodic function $G_{2\kappa }^+ (x)$ admits the Fourier expansion\\[-3mm]
\[
G_{2\kappa }^+ (x)=\sum_{\ell =-\infty }^\infty c_\ell \,  \mathrm{e}^{\frac{2\pi \mathrm{i}}{d}\ell x}\\[-1mm]
\]
with\\[-3mm]
\[
\begin{array}{rl}
c_\ell & =\frac{1}{d}\int _0^d\mathrm{e}^{-\frac{2\pi \mathrm{i}}{d}\ell x}\sum_{\alpha =-\infty }^\infty \mathrm{e}^{-\kappa \left(\sqrt{\frac{2\pi }{d}}\, \left(\left(\alpha + \frac{1}{2}\right)d +x\right)\right)^2} dx\\[2mm]
& =\frac{1}{d}\sum_{\alpha =-\infty }^\infty \int _0^d\mathrm{e}^{-\frac{2\pi \mathrm{i}}{d}\ell x} \mathrm{e}^{-\kappa \left(\sqrt{\frac{2\pi }{d}}\, \left(\left(\alpha + \frac{1}{2}\right)d +x\right)\right)^2} dx.
\\[-1mm]
\end{array}
\]
By denoting $t\!=\!\sqrt{\frac{2\pi }{d}}\, \left(\left(\alpha + \frac{1}{2}\right)d +x\right)$ 
we get\\[-3mm]
\[
\begin{array}{rl}
c_\ell & =\frac{1}{\sqrt{2\pi d}}\sum_{\alpha =-\infty }^\infty \int_{(\alpha -1/2)\sqrt{2\pi d}}^{(\alpha +1/2)\sqrt{2\pi d}}\mathrm{e}^{-\frac{2\pi \mathrm{i}}{d}\ell \left(t\sqrt{\frac{d}{2\pi }}-\left(\alpha +\frac{1}{2}\right) d\right)} \mathrm{e}^{-\kappa t^2}\, dt\\[4mm]
 & =\frac{(-1)^\ell }{\sqrt{2\pi d}}\sum_{\alpha =-\infty }^\infty \int_{(\alpha -1/2)\sqrt{2\pi d}}^{(\alpha +1/2)\sqrt{2\pi d}}\mathrm{e}^{-\mathrm{i}\ell  t\sqrt{\frac{2\pi }{d}}}
\mathrm{e}^{-\kappa t^2}\, dt\\[4mm]
& =\frac{(-1)^\ell }{\sqrt{2\pi d}}\int_{-\infty }^{\infty}\mathrm{e}^{-\mathrm{i}\ell  t\sqrt{\frac{2\pi }{d}}} \mathrm{e}^{-\kappa t^2}\,  dt=\frac{(-1)^\ell }{\sqrt{2\kappa d}}\, \mathrm{e}^{-\frac{\pi}{2\kappa d}\ell ^2}\\[-1mm]
\end{array}
\]
whence\\[-3mm]
\[
G_{2\kappa }^+(x)=\frac{1}{\sqrt{2\kappa d}}\,\sum_{\ell =-\infty }^\infty   \mathrm{e}^{\frac{2\pi \mathrm{i}}{d}\ell x}\,(-1)^\ell \,  \mathrm{e}^{-\frac{\pi}{2\kappa d}\ell ^2}.\\[-1mm]
\]
Particularly, we have\\[-3mm]
\[
\begin{array}{rl}
\mathfrak{g}_{2\kappa }^+(j) & =G_{2\kappa }^+(j)=\frac{1}{\sqrt{2\kappa d}}\,\sum_{\ell =-\infty }^\infty   \mathrm{e}^{\frac{2\pi \mathrm{i}}{d}j\ell }\,(-1)^\ell \, \mathrm{e}^{-\frac{\pi}{2\kappa d}\ell ^2}\\[3mm]
 & =\frac{1}{\sqrt{2\kappa d}}\sum_{n=-s}^{s}\sum_{\alpha =-\infty }^\infty  \mathrm{e}^{\frac{2\pi \mathrm{i}}{d}j(\alpha d+n) }\,(-1)^{(\alpha d+n)} \mathrm{e}^{-\frac{\pi }{2\kappa d}(\alpha d+n) ^2}\, \\[2mm]
& =\frac{1}{\sqrt{d}}\sum_{n=-s}^{s}\mathrm{e}^{\frac{2\pi \mathrm{i}}{d}jn }\frac{(-1)^n}{\sqrt{2\kappa }}\sum_{\alpha =-\infty }^\infty  \,(-1)^\alpha \,  \mathrm{e}^{-\frac{\pi }{2\kappa d}(\alpha d+n) ^2}\\[2mm]
& =\frac{1}{\sqrt{d}}\sum_{n=-s}^{s}\mathrm{e}^{-\frac{2\pi \mathrm{i}}{d}jn }\frac{(-1)^n}{\sqrt{2\kappa }}\sum_{\alpha =-\infty }^\infty  \,(-1)^\alpha \,  \mathrm{e}^{-\frac{\pi }{2\kappa d}(\alpha d+n) ^2}
\end{array}
\]
whence\\[-3mm] 
\[
\mathcal{F}[\mathfrak{g}_{2\kappa }^+](n)=\frac{(-1)^n}{\sqrt{2\kappa }}\sum_{\alpha =-\infty }^\infty  \,(-1)^\alpha \,  \mathrm{e}^{-\frac{\pi }{2\kappa d}(\alpha d+n) ^2}
\]
and we get
\[
\mathcal{F}[\mathfrak{g}_{2\kappa }^+](2m)\!=\!\frac{1}{\sqrt{2\kappa }}\!\sum_{\alpha =-\infty }^\infty  \!(-1)^\alpha \,  \mathrm{e}^{-\frac{\pi }{2\kappa d}(\alpha d+2m) ^2}\!=\!\frac{1}{\sqrt{2\kappa }}\left( \mathfrak{g}_{\frac{2}{\kappa }} (m)\!-\!\mathfrak{g}_{\frac{2}{\kappa }}^+ (m)\right).\quad \opensquare
\]
{\bf Lemma 2}. {\it If the numbers $N_{\alpha ,\beta }$ are such that the series are absolutely  convergent then} 
\[
\sum\limits_{\alpha ,\beta =-\infty }^\infty N_{\alpha ,\beta }=\sum\limits_{\mu  ,\eta =-\infty }^\infty N_{\mu +\eta ,\mu -\eta } + \sum\limits_{\mu  ,\eta  =-\infty }^\infty N_{\mu +\eta +1,\mu -\eta }.
\]
{\bf Proof}. We separate the sum as \cite{Marzoli}
\[
\sum\limits_{\alpha ,\beta =-\infty }^\infty N_{\alpha ,\beta }=\sum\limits_{\scriptsize 
\begin{array}{c}
\alpha ,\beta \\
{\rm both\ even}\\
{\rm or}\\
{\rm both\ odd}
\end{array}} N_{\alpha ,\beta }+\sum\limits_{\scriptsize 
\begin{array}{c}
\alpha ,\beta \\
{\rm one\ even}\\
{\rm and}\\
{\rm other\ odd}
\end{array}} N_{\alpha ,\beta }
\]
and use the substitutions $(\alpha , \beta )\!=\!(\mu \!+\!\eta ,\mu \!-\!\eta )$ and $(\alpha , \beta )\!=\!(\mu \!+\!\eta \!+\!1,\mu \!-\!\eta )$, respectively.\qquad \opensquare

 The function $W:\mathbb{Z}_d\times \mathbb{Z}_d\longrightarrow \mathbb{C}$
\begin{equation}\label{Wignerdef}
\begin{array}{rl} 
W(n,m) &\!\!\!\!=\frac{1}{d}\sum\limits_{k=-s}^s\mathrm{e}^{\frac{4\pi \mathrm{i}}{d}mk}\, \mathfrak{g}_\kappa (n-k)\, \mathfrak{g}_\kappa (n+k)\\[2mm]
 & \!\!\!\!=\frac{1}{d}\sum\limits_{k=-s}^s\mathrm{e}^{\frac{4\pi \mathrm{i}}{d}mk}\, 
\sum\limits_{\alpha ,\beta =-\infty }^\infty \mathrm{e}^{-\kappa  \frac{\pi }{d} (\alpha d +n-k)^2}
 \mathrm{e}^{-\kappa  \frac{\pi }{d} (\beta d +n+k)^2}
\end{array}
\end{equation}
is called the {\em discrete Wigner function} corresponding to $\mathfrak{g}_\kappa $. 
It is well-determined by its restriction to the unit cell
$\{ -s, -s\!+\!1,...,s\!-\!1,s\}\!\times \!\{ -s, -s\!+\!1,...,s\!-\!1,s\}$
directly related to the finite phase space $\mathcal{S}_d\!\times \! \mathcal{S}_d$. 
\begin{center}
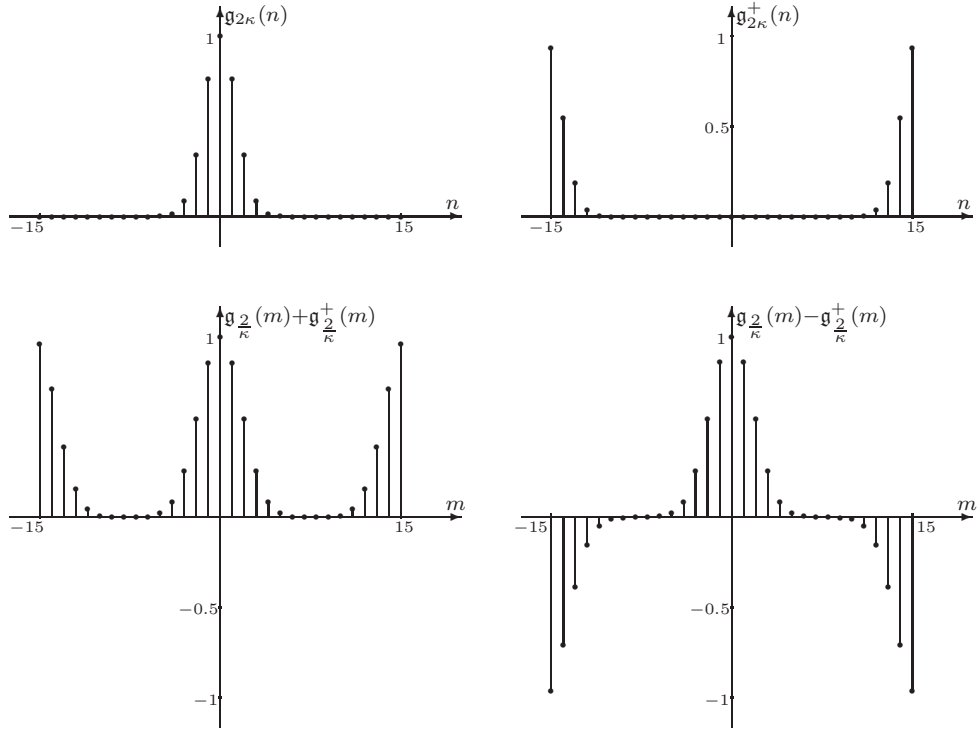
\begin{figure}[h]
\setlength{\unitlength}{2mm}
\begin{picture}(70,50)(-8,0)
\put(3,14){\vector(1,0){30}}
\put(37,14){\vector(1,0){30}}
\put(3,34){\vector(1,0){30}}
\put(37,34){\vector(1,0){30}}

\put(17,0){\vector(0,1){28}}
\put(51,0){\vector(0,1){28}}
\put(17,32){\vector(0,1){16}}
\put(51,32){\vector(0,1){16}}

\put(    5.00000,  34.00000){\circle*{0.3}}
\put(    5.80000,  34.00000){\circle*{0.3}}
\put(    6.60000,  34.00000){\circle*{0.3}}
\put(    7.40000,  34.00000){\circle*{0.3}}
\put(    8.20000,  34.00000){\circle*{0.3}}
\put(    9.00000,  34.00000){\circle*{0.3}}
\put(    9.80000,  34.00000){\circle*{0.3}}
\put(   10.60000,  34.00000){\circle*{0.3}}
\put(   11.40000,  34.00000){\circle*{0.3}}
\put(   12.20000,  34.00071){\circle*{0.3}}
\put(   13.00000,  34.01397){\circle*{0.3}}
\put(   13.80000,  34.15897){\circle*{0.3}}
\put(   14.60000,  35.05412){\circle*{0.3}}
\put(   15.40000,  38.07117){\circle*{0.3}}
\put(   16.20000,  43.15832){\circle*{0.3}}
\put(   17.00000,  46.00000){\circle*{0.3}}
\put(   17.80000,  43.15832){\circle*{0.3}}
\put(   18.60000,  38.07117){\circle*{0.3}}
\put(   19.40000,  35.05412){\circle*{0.3}}
\put(   20.20000,  34.15897){\circle*{0.3}}
\put(   21.00000,  34.01397){\circle*{0.3}}
\put(   21.80000,  34.00071){\circle*{0.3}}
\put(   22.60000,  34.00000){\circle*{0.3}}
\put(   23.40000,  34.00000){\circle*{0.3}}
\put(   24.20000,  34.00000){\circle*{0.3}}
\put(   25.00000,  34.00000){\circle*{0.3}}
\put(   25.80000,  34.00000){\circle*{0.3}}
\put(   26.60000,  34.00000){\circle*{0.3}}
\put(   27.40000,  34.00000){\circle*{0.3}}
\put(   28.20000,  34.00000){\circle*{0.3}}
\put(   29.00000,  34.00000){\circle*{0.3}}
\put(    5.00000,34){\line(0,1){    .00000}}
\put(    5.80000,34){\line(0,1){    .00000}}
\put(    6.60000,34){\line(0,1){    .00000}}
\put(    7.40000,34){\line(0,1){    .00000}}
\put(    8.20000,34){\line(0,1){    .00000}}
\put(    9.00000,34){\line(0,1){    .00000}}
\put(    9.80000,34){\line(0,1){    .00000}}
\put(   10.60000,34){\line(0,1){    .00000}}
\put(   11.40000,34){\line(0,1){    .00000}}
\put(   12.20000,34){\line(0,1){    .00071}}
\put(   13.00000,34){\line(0,1){    .01396}}
\put(   13.80000,34){\line(0,1){    .15897}}
\put(   14.60000,34){\line(0,1){   1.05412}}
\put(   15.40000,34){\line(0,1){   4.07117}}
\put(   16.20000,34){\line(0,1){   9.15832}}
\put(   17.00000,34){\line(0,1){  12.00000}}
\put(   17.80000,34){\line(0,1){   9.15832}}
\put(   18.60000,34){\line(0,1){   4.07117}}
\put(   19.40000,34){\line(0,1){   1.05412}}
\put(   20.20000,34){\line(0,1){    .15897}}
\put(   21.00000,34){\line(0,1){    .01396}}
\put(   21.80000,34){\line(0,1){    .00071}}
\put(   22.60000,34){\line(0,1){    .00000}}
\put(   23.40000,34){\line(0,1){    .00000}}
\put(   24.20000,34){\line(0,1){    .00000}}
\put(   25.00000,34){\line(0,1){    .00000}}
\put(   25.80000,34){\line(0,1){    .00000}}
\put(   26.60000,34){\line(0,1){    .00000}}
\put(   27.40000,34){\line(0,1){    .00000}}
\put(   28.20000,34){\line(0,1){    .00000}}
\put(   29.00000,34){\line(0,1){    .00000}}
\put(   39.00000,  45.21605){\circle*{0.3}}
\put(   39.80000,  40.53293){\circle*{0.3}}
\put(   40.60000,  36.21639){\circle*{0.3}}
\put(   41.40000,  34.43798){\circle*{0.3}}
\put(   42.20000,  34.05041){\circle*{0.3}}
\put(   43.00000,  34.00338){\circle*{0.3}}
\put(   43.80000,  34.00013){\circle*{0.3}}
\put(   44.60000,  34.00000){\circle*{0.3}}
\put(   45.40000,  34.00000){\circle*{0.3}}
\put(   46.20000,  34.00000){\circle*{0.3}}
\put(   47.00000,  34.00000){\circle*{0.3}}
\put(   47.80000,  34.00000){\circle*{0.3}}
\put(   48.60000,  34.00000){\circle*{0.3}}
\put(   49.40000,  34.00000){\circle*{0.3}}
\put(   50.20000,  34.00000){\circle*{0.3}}
\put(   51.00000,  34.00000){\circle*{0.3}}
\put(   51.80000,  34.00000){\circle*{0.3}}
\put(   52.60000,  34.00000){\circle*{0.3}}
\put(   53.40000,  34.00000){\circle*{0.3}}
\put(   54.20000,  34.00000){\circle*{0.3}}
\put(   55.00000,  34.00000){\circle*{0.3}}
\put(   55.80000,  34.00000){\circle*{0.3}}
\put(   56.60000,  34.00000){\circle*{0.3}}
\put(   57.40000,  34.00000){\circle*{0.3}}
\put(   58.20000,  34.00013){\circle*{0.3}}
\put(   59.00000,  34.00338){\circle*{0.3}}
\put(   59.80000,  34.05041){\circle*{0.3}}
\put(   60.60000,  34.43798){\circle*{0.3}}
\put(   61.40000,  36.21639){\circle*{0.3}}
\put(   62.20000,  40.53293){\circle*{0.3}}
\put(   63.00000,  45.21605){\circle*{0.3}}
\put(   39.00000,34){\line(0,1){  11.21605}}
\put(   39.80000,34){\line(0,1){   6.53293}}
\put(   40.60000,34){\line(0,1){   2.21639}}
\put(   41.40000,34){\line(0,1){    .43798}}
\put(   42.20000,34){\line(0,1){    .05041}}
\put(   43.00000,34){\line(0,1){    .00338}}
\put(   43.80000,34){\line(0,1){    .00013}}
\put(   44.60000,34){\line(0,1){    .00000}}
\put(   45.40000,34){\line(0,1){    .00000}}
\put(   46.20000,34){\line(0,1){    .00000}}
\put(   47.00000,34){\line(0,1){    .00000}}
\put(   47.80000,34){\line(0,1){    .00000}}
\put(   48.60000,34){\line(0,1){    .00000}}
\put(   49.40000,34){\line(0,1){    .00000}}
\put(   50.20000,34){\line(0,1){    .00000}}
\put(   51.00000,34){\line(0,1){    .00000}}
\put(   51.80000,34){\line(0,1){    .00000}}
\put(   52.60000,34){\line(0,1){    .00000}}
\put(   53.40000,34){\line(0,1){    .00000}}
\put(   54.20000,34){\line(0,1){    .00000}}
\put(   55.00000,34){\line(0,1){    .00000}}
\put(   55.80000,34){\line(0,1){    .00000}}
\put(   56.60000,34){\line(0,1){    .00000}}
\put(   57.40000,34){\line(0,1){    .00000}}
\put(   58.20000,34){\line(0,1){    .00013}}
\put(   59.00000,34){\line(0,1){    .00338}}
\put(   59.80000,34){\line(0,1){    .05041}}
\put(   60.60000,34){\line(0,1){    .43798}}
\put(   61.40000,34){\line(0,1){   2.21639}}
\put(   62.20000,34){\line(0,1){   6.53293}}
\put(   63.00000,34){\line(0,1){  11.21605}}
\put(    5.00000,  25.55252){\circle*{0.3}}
\put(    5.80000,  22.52394){\circle*{0.3}}
\put(    6.60000,  18.64053){\circle*{0.3}}
\put(    7.40000,  15.86406){\circle*{0.3}}
\put(    8.20000,  14.55247){\circle*{0.3}}
\put(    9.00000,  14.12082){\circle*{0.3}}
\put(    9.80000,  14.01955){\circle*{0.3}}
\put(   10.60000,  14.00304){\circle*{0.3}}
\put(   11.40000,  14.00719){\circle*{0.3}}
\put(   12.20000,  14.05042){\circle*{0.3}}
\put(   13.00000,  14.26837){\circle*{0.3}}
\put(   13.80000,  15.05412){\circle*{0.3}}
\put(   14.60000,  17.05504){\circle*{0.3}}
\put(   15.40000,  20.53293){\circle*{0.3}}
\put(   16.20000,  24.30774){\circle*{0.3}}
\put(   17.00000,  26.00000){\circle*{0.3}}
\put(   17.80000,  24.30774){\circle*{0.3}}
\put(   18.60000,  20.53293){\circle*{0.3}}
\put(   19.40000,  17.05504){\circle*{0.3}}
\put(   20.20000,  15.05412){\circle*{0.3}}
\put(   21.00000,  14.26837){\circle*{0.3}}
\put(   21.80000,  14.05042){\circle*{0.3}}
\put(   22.60000,  14.00719){\circle*{0.3}}
\put(   23.40000,  14.00304){\circle*{0.3}}
\put(   24.20000,  14.01955){\circle*{0.3}}
\put(   25.00000,  14.12082){\circle*{0.3}}
\put(   25.80000,  14.55247){\circle*{0.3}}
\put(   26.60000,  15.86406){\circle*{0.3}}
\put(   27.40000,  18.64053){\circle*{0.3}}
\put(   28.20000,  22.52394){\circle*{0.3}}
\put(   29.00000,  25.55252){\circle*{0.3}}
\put(    5.00000,14){\line(0,1){  11.55252}}
\put(    5.80000,14){\line(0,1){   8.52394}}
\put(    6.60000,14){\line(0,1){   4.64053}}
\put(    7.40000,14){\line(0,1){   1.86406}}
\put(    8.20000,14){\line(0,1){    .55248}}
\put(    9.00000,14){\line(0,1){    .12082}}
\put(    9.80000,14){\line(0,1){    .01955}}
\put(   10.60000,14){\line(0,1){    .00304}}
\put(   11.40000,14){\line(0,1){    .00719}}
\put(   12.20000,14){\line(0,1){    .05042}}
\put(   13.00000,14){\line(0,1){    .26837}}
\put(   13.80000,14){\line(0,1){   1.05412}}
\put(   14.60000,14){\line(0,1){   3.05504}}
\put(   15.40000,14){\line(0,1){   6.53293}}
\put(   16.20000,14){\line(0,1){  10.30774}}
\put(   17.00000,14){\line(0,1){  12.00000}}
\put(   17.80000,14){\line(0,1){  10.30774}}
\put(   18.60000,14){\line(0,1){   6.53293}}
\put(   19.40000,14){\line(0,1){   3.05504}}
\put(   20.20000,14){\line(0,1){   1.05412}}
\put(   21.00000,14){\line(0,1){    .26837}}
\put(   21.80000,14){\line(0,1){    .05042}}
\put(   22.60000,14){\line(0,1){    .00719}}
\put(   23.40000,14){\line(0,1){    .00304}}
\put(   24.20000,14){\line(0,1){    .01955}}
\put(   25.00000,14){\line(0,1){    .12082}}
\put(   25.80000,14){\line(0,1){    .55248}}
\put(   26.60000,14){\line(0,1){   1.86406}}
\put(   27.40000,14){\line(0,1){   4.64053}}
\put(   28.20000,14){\line(0,1){   8.52394}}
\put(   29.00000,14){\line(0,1){  11.55252}}
\put(   39.00000,   2.44748){\circle*{0.3}}
\put(   39.80000,   5.47606){\circle*{0.3}}
\put(   40.60000,   9.35947){\circle*{0.3}}
\put(   41.40000,  12.13594){\circle*{0.3}}
\put(   42.20000,  13.44753){\circle*{0.3}}
\put(   43.00000,  13.87919){\circle*{0.3}}
\put(   43.80000,  13.98056){\circle*{0.3}}
\put(   44.60000,  13.99839){\circle*{0.3}}
\put(   45.40000,  14.00678){\circle*{0.3}}
\put(   46.20000,  14.05040){\circle*{0.3}}
\put(   47.00000,  14.26836){\circle*{0.3}}
\put(   47.80000,  15.05412){\circle*{0.3}}
\put(   48.60000,  17.05504){\circle*{0.3}}
\put(   49.40000,  20.53293){\circle*{0.3}}
\put(   50.20000,  24.30774){\circle*{0.3}}
\put(   51.00000,  26.00000){\circle*{0.3}}
\put(   51.80000,  24.30774){\circle*{0.3}}
\put(   52.60000,  20.53293){\circle*{0.3}}
\put(   53.40000,  17.05504){\circle*{0.3}}
\put(   54.20000,  15.05412){\circle*{0.3}}
\put(   55.00000,  14.26836){\circle*{0.3}}
\put(   55.80000,  14.05040){\circle*{0.3}}
\put(   56.60000,  14.00678){\circle*{0.3}}
\put(   57.40000,  13.99839){\circle*{0.3}}
\put(   58.20000,  13.98056){\circle*{0.3}}
\put(   59.00000,  13.87919){\circle*{0.3}}
\put(   59.80000,  13.44753){\circle*{0.3}}
\put(   60.60000,  12.13594){\circle*{0.3}}
\put(   61.40000,   9.35947){\circle*{0.3}}
\put(   62.20000,   5.47606){\circle*{0.3}}
\put(   63.00000,   2.44748){\circle*{0.3}}
\put(   39.00000,14){\line(0,-1){ 11.55252}}
\put(   39.80000,14){\line(0,-1){  8.52394}}
\put(   40.60000,14){\line(0,-1){  4.64053}}
\put(   41.40000,14){\line(0,-1){  1.86406}}
\put(   42.20000,14){\line(0,-1){   .55248}}
\put(   43.00000,14){\line(0,-1){   .12081}}
\put(   43.80000,14){\line(0,-1){   .01944}}
\put(   44.60000,14){\line(0,-1){   .00161}}
\put(   45.40000,14){\line(0,1){    .00678}}
\put(   46.20000,14){\line(0,1){    .05040}}
\put(   47.00000,14){\line(0,1){    .26836}}
\put(   47.80000,14){\line(0,1){   1.05412}}
\put(   48.60000,14){\line(0,1){   3.05504}}
\put(   49.40000,14){\line(0,1){   6.53293}}
\put(   50.20000,14){\line(0,1){  10.30774}}
\put(   51.00000,14){\line(0,1){  12.00000}}
\put(   51.80000,14){\line(0,1){  10.30774}}
\put(   52.60000,14){\line(0,1){   6.53293}}
\put(   53.40000,14){\line(0,1){   3.05504}}
\put(   54.20000,14){\line(0,1){   1.05412}}
\put(   55.00000,14){\line(0,1){    .26836}}
\put(   55.80000,14){\line(0,1){    .05040}}
\put(   56.60000,14){\line(0,1){    .00678}}
\put(   57.40000,14){\line(0,-1){   .00161}}
\put(   58.20000,14){\line(0,-1){   .01944}}
\put(   59.00000,14){\line(0,-1){   .12081}}
\put(   59.80000,14){\line(0,-1){   .55248}}
\put(   60.60000,14){\line(0,-1){  1.86406}}
\put(   61.40000,14){\line(0,-1){  4.64053}}
\put(   62.20000,14){\line(0,-1){  8.52394}}
\put(   63.00000,14){\line(0,-1){ 11.55252}}

\put(3,13){$\scriptscriptstyle{-15}$}
\put(5, 13.9){\line(0,1){0.3}}
\put(28.6,13){$\scriptscriptstyle{15}$}
\put(29, 13.9){\line(0,1){0.3}}
\put(3,33){$\scriptscriptstyle{-15}$}
\put(5, 33.9){\line(0,1){0.3}}
\put(28.6,33){$\scriptscriptstyle{15}$}
\put(29, 33.9){\line(0,1){0.3}}

\put(16,25.5){$\scriptscriptstyle{1}$}
\put(16.9, 26){\line(1,0){0.2}}
\put(15.2,1.5){$\scriptscriptstyle{-1}$}
\put(16.9, 2){\line(1,0){0.2}}
\put(14.2,7.55){$\scriptscriptstyle{-0.5}$}
\put(16.9, 8){\line(1,0){0.2}}
\put(16,45.5){$\scriptscriptstyle{1}$}
\put(16.9, 46){\line(1,0){0.2}}
%\put(14.3,39.55){$\scriptscriptstyle{0.5}$}
%\put(16.9, 40){\line(1,0){0.2}}

\put(32,14.5){$\scriptstyle{m}$}
\put(32,34.5){$\scriptstyle{n}$}
\put(17.3,27){$\scriptstyle \mathfrak{g}_{\frac{2}{\kappa } }(m)+\mathfrak{g}_{\frac{2}{\kappa } }^+(m)$}
\put(17.3,47){$\scriptstyle \mathfrak{g}_{2\kappa  }(n)$}

\put(36.5,13){$\scriptscriptstyle{-15}$}
\put(39, 13.9){\line(0,1){0.3}}
\put(63.3,13){$\scriptscriptstyle{15}$}
\put(63, 13.9){\line(0,1){0.3}}
\put(37.5,33){$\scriptscriptstyle{-15}$}
\put(39, 33.9){\line(0,1){0.3}}
\put(62.6,33){$\scriptscriptstyle{15}$}
\put(63, 33.9){\line(0,1){0.3}}

\put(50,25.5){$\scriptscriptstyle{1}$}
\put(50.9, 26){\line(1,0){0.2}}
\put(49.2,1.5){$\scriptscriptstyle{-1}$}
\put(50.9, 2){\line(1,0){0.2}}
\put(48.2,7.55){$\scriptscriptstyle{-0.5}$}
\put(50.9, 8){\line(1,0){0.2}}
\put(50,45.5){$\scriptscriptstyle{1}$}
\put(50.9, 46){\line(1,0){0.2}}
\put(49.2,39.55){$\scriptscriptstyle{0.5}$}
\put(50.9, 40){\line(1,0){0.2}}

\put(66,14.5){$\scriptstyle{m}$}
\put(66,34.5){$\scriptstyle{n}$}
\put(51.3,27){$\scriptstyle{\mathfrak{g}_{\frac{2}{\kappa }} (m)-\mathfrak{g}_{\frac{2}{\kappa }}^+ (m)}$}
\put(51.3,47){$\scriptstyle{\mathfrak{g}_{2 \kappa }^+ (n)}$}

\end{picture}
\caption{The functions involved in relation (\ref{wignergk})  in the case $\kappa =\frac{4}{3}$, $d\!=\!31$.}\label{Wigner}
\end{figure}
\end{center}
{\bf Theorem 4}. {\it The discrete Wigner function $W$ is a sum of products of finite Gaussians}
\begin{equation}\label{wignergk}
\begin{array}{rl} 
W(n,m) & =\frac{1}{\sqrt{2\kappa d}}\, \,  \mathfrak{g}_{2\kappa }(n)\, \left( \mathfrak{g}_{\frac{2}{\kappa } }(m)+\mathfrak{g}_{\frac{2}{\kappa } }^+(m)\right)\\[3mm]
& \ \ +\frac{1}{\sqrt{2\kappa d}}\,  \mathfrak{g}_{2\kappa }^+(n)\,\left( \mathfrak{g}_{\frac{2}{\kappa }} (m)-\mathfrak{g}_{\frac{2}{\kappa }}^+ (m)\right).
\end{array}
\end{equation}
{\bf Proof}. By using theorem 1, lemma 1 and lemma 2 we get
\[ \fl
\begin{array}{rl}
W(n,m)&=\frac{1}{d}\sum\limits_{k=-s}^s\mathrm{e}^{\frac{4\pi \mathrm{i}}{d}mk}\, 
\sum\limits_{\mu  ,\eta  =-\infty }^\infty \mathrm{e}^{-\kappa  \frac{\pi }{d} ((\mu +\eta ) d +n-k)^2}\ 
 \mathrm{e}^{-\kappa  \frac{\pi }{d} ((\mu -\eta ) d +n+k)^2}\\[4mm]
& \, \  +\frac{1}{d}\sum\limits_{k=-s}^s\mathrm{e}^{\frac{4\pi \mathrm{i}}{d}mk}\, 
\sum\limits_{\mu  ,\eta  =-\infty }^\infty \mathrm{e}^{-\kappa  \frac{\pi }{d} ((\mu +\eta +1) d +n-k)^2}\ 
 \mathrm{e}^{-\kappa  \frac{\pi }{d} ((\mu -\eta ) d +n+k)^2}\\[4mm]
& =\frac{1}{d}\sum\limits_{k=-s}^s\mathrm{e}^{\frac{4\pi \mathrm{i}}{d}mk}\, 
\sum\limits_{\mu  ,\eta  =-\infty }^\infty \mathrm{e}^{-2\kappa  \frac{\pi }{d} (\mu d +n)^2}\ 
 \mathrm{e}^{-2\kappa  \frac{\pi }{d} (\eta  d -k)^2}\\[4mm]
& \, \ +\frac{1}{d}\sum\limits_{k=-s}^s\mathrm{e}^{\frac{4\pi \mathrm{i}}{d}mk}\, 
\sum\limits_{\mu  ,\eta  =-\infty }^\infty \mathrm{e}^{-2\kappa  \frac{\pi }{d} \left(\left(\mu +\frac{1}{2}\right)d +n\right)^2}\ 
 \mathrm{e}^{-2\kappa  \frac{\pi }{d} \left(\left(\eta +\frac{1}{2}\right) d -k\right)^2}\\[4mm]
\end{array}
\]
\[ \fl
\begin{array}{rl}
\mbox{}\qquad & =\frac{1}{\sqrt{d}}\sum\limits_{\mu   =-\infty }^\infty \mathrm{e}^{-2\kappa  \frac{\pi }{d} (\mu d +n)^2}\ \frac{1}{\sqrt{d}}\sum\limits_{k=-s}^s\mathrm{e}^{\frac{4\pi \mathrm{i}}{d}mk}\, \sum\limits_{\eta  =-\infty }^\infty  \mathrm{e}^{-2\kappa  \frac{\pi }{d} (\eta  d -k)^2}\\[4mm]
& \, \ +\frac{1}{\sqrt{d}}\sum\limits_{\mu    =-\infty }^\infty \mathrm{e}^{-2\kappa  \frac{\pi }{d} \left(\left(\mu +\frac{1}{2}\right)d +n\right)^2}\ \frac{1}{\sqrt{d}}\sum\limits_{k=-s}^s\mathrm{e}^{\frac{4\pi \mathrm{i}}{d}mk}\, 
\sum\limits_{\eta  =-\infty }^\infty 
 \mathrm{e}^{-2\kappa  \frac{\pi }{d} \left(\left(\eta +\frac{1}{2}\right) d -k\right)^2}\\[4mm]
& =\frac{1}{\sqrt{d}}\, \,  \mathfrak{g}_{2\kappa }(n)\, \, \mathcal{F}[\mathfrak{g}_{2\kappa }](2m)
 +\frac{1}{\sqrt{d}}\, \, \mathfrak{g}_{2\kappa }^+(n)\,\, \ \frac{1}{\sqrt{d}}\sum\limits_{k=-s}^s\mathrm{e}^{\frac{4\pi \mathrm{i}}{d}mk}\, 
\sum\limits_{\eta  =-\infty }^\infty 
 \mathrm{e}^{-2\kappa  \frac{\pi }{d} \left(\left(\eta -\frac{1}{2}\right) d +k\right)^2}\\[4mm]
& =\frac{1}{\sqrt{2\kappa d}}\, \,  \mathfrak{g}_{2\kappa }(n)\, \, \mathfrak{g}_{\frac{1}{2\kappa } }(2m)
 +\frac{1}{\sqrt{d}}\, \, \mathfrak{g}_{2\kappa }^+(n)\,\,  \mathcal{F}[\mathfrak{g}_{2\kappa }^+](2m)\\[4mm]
& =\frac{1}{\sqrt{2\kappa d}}\, \,  \mathfrak{g}_{2\kappa }(n)\, \left( \mathfrak{g}_{\frac{2}{\kappa } }(m)+
\mathfrak{g}_{\frac{2}{\kappa } }^+(m)\right)+\frac{1}{\sqrt{2\kappa d}}\,  \mathfrak{g}_{2\kappa }^+(n)\,\left( \mathfrak{g}_{\frac{2}{\kappa }} (m)-\mathfrak{g}_{\frac{2}{\kappa }}^+ (m)\right).\qquad \opensquare
\end{array}
\]

\begin{figure}
\centering
\includegraphics[scale=0.7]{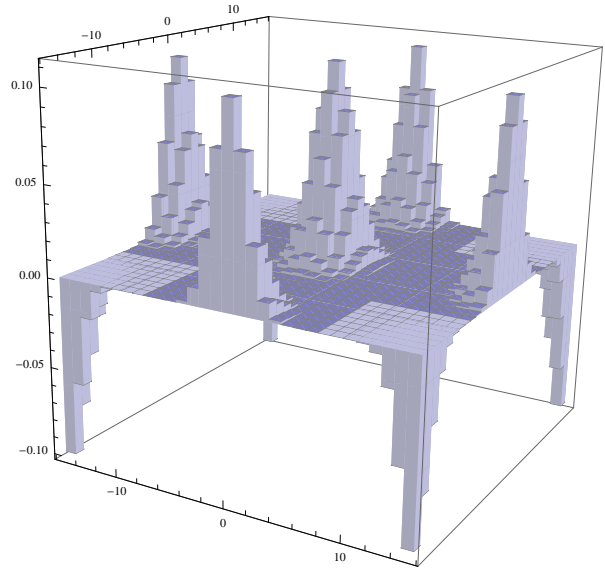}
\caption{\label{fig4} The Wigner function $W(n,m)$ in the case $\kappa =\frac{4}{3}$, $d\!=\!31$. }
\end{figure}

\noindent 
The shape of the functions involved in the expression (\ref{wignergk}) of $W$ can be seen in  Figure \ref{Wigner}. In the finite phase space (which is a finite torus), the obtained discrete Wigner function $W$ has three peaks placed  around $(0,0)$, $(0,s)$, $(s,0)$ and an anti-peak around $(s,s)$ (see Figure \ref{fig4}).  Note the similarity of the representation in Figure \ref{fig4} and the results in \cite{marchioli,Opatrny1,Opatrny2}.

The Wigner function from the continuous case is, in a certain sense, the limit of the discrete Wigner function. Therefore, in the continuous limit, the only``surviving'' peak is expected to be that around $(0,0)$, in agreement with the known results from the Wigner function of continuous Gaussians.
The disappearance of the peaks placed  around $(0,s)$, $(s,0)$ and of the anti-peak around $(s,s)$
in the continuous limit is rather mysterious. The numerical simulations show that for large $d$ all of them keep an amplitude comparable to that of the peak placed  around $(0,0)$. A possible explanation is the following. In finite dimension these peaks are placed on the ``boundary'' of the finite phase space $\mathcal{S}_d\times \mathcal{S}_d$. In the continuous limit they must be located on the ``boundary'' of the phase plane $\mathbb{R}^2$ which contains just one point, namely $\infty $.
It is known that the compactification of the real or complex plane is obtained by adding one point, and the extended plane (called Riemann sphere) corresponds to the unit sphere through the stereographic projection. We can consider that, for $d\longrightarrow \infty $, that is, for $s\longrightarrow \infty $,  the peaks placed  around $(0,s)$, $(s,0)$
compensate with the anti-peak around $(s,s)$.
\bigskip

\section{Discussions and Conclusions}

The simpler way to define a Gaussian type function on the set $\{ -s,-s\!+\!1,\, ...\, ,s\!-\!1,s\}$ or $\mathcal{S}_d$ is to consider the restriction of the Gaussian $g_\kappa (x)\!=\!\mathrm{e}^{-\frac{\kappa }{2}x^2}$ to these sets
\[
f:\{ -s,-s\!+\!1,\, ...\, ,s\!-\!1,s\}\longrightarrow \mathbb{R}, \qquad  f(n)={\rm e}^{-\frac{\kappa }{2}n^2}
\]
respectively, 
\[
\begin{array}{l}
f:\mathcal{S}_d\longrightarrow \mathbb{R}, \qquad  f\left( n\sqrt{\frac{2\pi }{d}}\right)= {\rm e}^{-\kappa \frac{\pi }{d}n^2}.
\end{array}
\]
The functions obtained in this way behave similar to the continuous  Gaussians only for large values of $d\!=\!2s\!+\!1$.

Our approach is based on an alternative method. By using a Weil-Zak type transform, we firstly generate a periodic function $G_\kappa $ with period $d$, and then we define our finite Gaussian as a restriction of $G_\kappa $ to $\mathbb{Z}$, namely $\mathfrak{g}(n)\!=\!G_\kappa (n)$.
Our finite Gaussians starts to behave similar to a continuous Gaussian from relatively small values of $d$. In addition, they have several  remarkable mathematical properties.

The discreteness of the configuration space has a particular appeal in both classical and quantum physics, particularly in phase space \cite{poletti} or in attempts to unify general relativity and quantum mechanics \cite{marchioli}. On the other hand, the results in a discrete configuration space are required to match standard physical results in the continuum limit. Because discrete physical systems are of interest in quantum mechanics or in classical physics in connection to the problem of sampling, the characterization of such systems and of their evolution has already received some attention. For example, a discrete quantum phase space has been studied in \cite{santhanam1,santhanam2}, the continuous limit being obtained by decreasing the lattice spacing in both momentum and position space, while in \cite{Bang2006} the dynamics on a discrete quantum phase space has been investigated by defining a Hamiltonian with the appropriate classical limit. The results in this paper are different from previous ones and demonstrate the importance of the choice of some  particular states  in the discrete-continuum transition. As an additional example in this respect, we mention that a discrete model of the quantum harmonic oscillator, for instance, can be introduced such that the energy spectrum is equally-spaced and the spectra of both momentum and position operators are denumerable non-degenerate \cite{wolf}. This last model, too, recovers the results for the ordinary harmonic oscillator in an appropriate limit. As do the discrete models of the quantum harmonic oscillator in terms of Kravchuk polynomials \cite{lorente} or Harper functions \cite{barker}. 

Summarizing, the correct continuous limit of discrete models can be obtained in many situations. The choice of Gaussian states render the discrete-continuous transition smoother in the sense that many results known from the continuous case are obtained for reasonably large d values or can easily be extrapolated from the results for finite d values. Moreover, Gaussian states have advantages over other states in phase space representations of physical systems since, as shown in Section 7, the Wigner distribution function has a particularly simple expression in this case. In conclusion, finite Gaussian states represent a useful mathematical tool for the study of quantum or classical physical systems. 

\end{document}